\documentclass[%
reprint,
superscriptaddress,
showpacs,preprintnumbers,
 amsmath,amssymb,
 aps,
pra,
dvipdfmx,
]{revtex4-1}

\usepackage[dvipdfmx]{graphicx}% Include figure files
\usepackage{dcolumn}% Align table columns on decimal point
\usepackage{bm}% bold math
\usepackage[dvipdfmx,
colorlinks=true,
linkcolor=blue,
filecolor=blue,
citecolor=blue,
urlcolor=blue
]{hyperref}% add hypertext capabilities
\usepackage{comment}

\begin{document}

%\preprint{APS/123-QED}

\title{Molecular adsorbed layer formation on cooled mirrors and its impacts on cryogenic gravitational wave telescopes}

\author{Kunihiko Hasegawa}
\email{knhkhsgw@icrr.u-tokyo.ac.jp}
\affiliation{Institute for Cosmic Ray Research,  The University of Tokyo, Kashiwa, Chiba 277-8582, Japan}

\author{Tomotada Akutsu}
\affiliation{National Astronomical Observatory of Japan, Mitaka, Tokyo 181-8588, Japan}

\author{Nobuhiro Kimura}
\affiliation{High Energy Accelerator Research Organization, Tsukuba, Ibaraki 305-0801, Japan}
\affiliation{The Graduate University for Advanced Studies, Tsukuba, Ibaraki, 305-0801, Japan}

\author{Yoshio Saito}
\affiliation{Institute for Cosmic Ray Research, The University of Tokyo, Kashiwa, Chiba 277-8582, Japan}

\author{Toshikazu Suzuki}
\affiliation{Institute for Cosmic Ray Research, The University of Tokyo, Kashiwa, Chiba 277-8582, Japan}
\affiliation{High Energy Accelerator Research Organization, Tsukuba, Ibaraki 305-0801, Japan}

\author{Takayuki Tomaru}
\affiliation{High Energy Accelerator Research Organization, Tsukuba, Ibaraki 305-0801, Japan}
\affiliation{The Graduate University for Advanced Studies, Tsukuba, Ibaraki, 305-0801, Japan}

\author{Ayako Ueda}
\affiliation{High Energy Accelerator Research Organization, Tsukuba, Ibaraki 305-0801, Japan}

\author{Shinji Miyoki}
\email{miyoki@icrr.u-tokyo.ac.jp}
\affiliation{Institute for Cosmic Ray Research, The University of Tokyo, Kashiwa, Chiba 277-8582, Japan}

\date{\today}% It is always \today, today,

%%%%%%%%%%%%%%%%%%%%%%%%%%%%%%%%%%%%%%%%%%%%%%%%%%%%%%%%%%%
\begin{abstract}
 Cryogenic mirrors have been introduced to the KAGRA gravitational wave telescope in Japan, and are also planned to be used in next-generation gravitational wave telescopes to further improve their sensitivity. Molecular gases inside vacuum chambers adhere to cold mirror surfaces because they lose their kinetic energy when they hit cryogenic surfaces. Finally, a number of adsorbed molecules form an adlayer, which will grow with time. The growing adlayer functions as an optical coating and changes the properties of the underlying mirror, such as reflectance, transmittance, and absorption, which are carefully chosen to maximize the detector sensitivity. The adlayer possibly affects the gravitational wave detector sensitivity. In order to characterize these changes, a high-finesse Fabry--Perot cavity was introduced to a KAGRA cryostat and the finesse of the cavity was monitored for 35 days under cryogenic conditions. We confirmed that the molecular adlayer was formed on a cold mirror and caused an oscillation in the finesse. The real and imaginary parts of the refractive index of the adlayer were $1.26 \pm 0.073$ and $2.2 \times 10^{-7} \pm 1.3 \times 10^{-7} $, respectively. These are considered to be that of $\mathrm{H_2O}$ molecules. The formation rate of the molecular adlayer was 27 $\pm$ 1.9 $\mathrm{nm/day}$. In this paper, we describe theoretical and experimental studies of the formation of a molecular adlayer on cryogenic mirrors. Furthermore, the effects of a molecular adlayer on the quantum noise and the input heat to the test mass are also discussed.
%\begin{description}
%\end{description}
\end{abstract}

%\pacs{03.65.Ta, 04.80.Nn, 42.60.Da}
\maketitle

%\tableofcontents

%%%%%%%%%%%%%%%%%%%%%%%%%%%%%%%%%%%%%%%%%%%%%%%%%%%%%%%%%%%
\section{Introduction}
Gravitational waves (GWs) from binary black hole mergers were detected by the Advanced Laser Interferometric Gravitational wave Observatory (LIGO) and Advanced Virgo \cite{Abbott2016}. However, higher sensitivities are essential to reveal the nature of GWs \cite{Abbott2017a}, the evolution of black holes \cite{Kinugawa2014}, the Hubble constant. The sensitivity of an interferometric gravitational wave detector (GWD) is mainly limited by certain noises, such as photon radiation pressure noise, photon shot noise, and thermal noise. In particular, mirror and suspension thermal noise limit the most sensitive frequency band from 10 to 100 Hz \cite{Harry2007}. In order to achieve higher sensitivity, the improvement of thermal noise is essential and certain solutions have been proposed, such as the introduction of larger mirrors, coating films that have high mechanical quality factors \cite{Cole2013}, and a Fabry--Perot (FP) arm cavity that resonates higher-order modes inside \cite{Mours2006} for instance. Additionally, the introduction of cryogenic mirrors and suspensions is also one of the ways to reduce thermal noise. In contrast to LIGO and Virgo, KAGRA \cite{Somiya2012,Aso2013} introduces cryogenic techniques for this purpose by using sapphire substrates for test masses and sapphire fibers for suspension wires because of its advantage of thermal conductivity and mechanical Q factor at cryogenic temperature \cite{Khalaidovski2014,Uchiyama1999}. Silicon also has these features \cite{Amico2004,Nawrodt2008} and LIGO plans to introduce 120 K silicon mirrors in their future upgrade \cite{Amico2004}, and Einstein Telescope (ET), the European next-generation GWD, also plans to introduce cryogenic mirrors \cite{Punturo2010,Hild2012}.\\
\quad In order to practically introduce cryogenic techniques for the thermal noise reduction in KAGRA, several properties of sapphire have been measured and tested at cryogenic temperatures as one of the candidates of the material for mirrors and fibers \cite{Tomaru2001,Uchiyama1998}. Ultralow-vibration cryocoolers \cite{Tomaru2004,Tokoku2014}, heat radiation shield ducts with baffles \cite{Sakakibara2015}, and pure aluminum heat link wires were also developed. \\
\quad The mirror reflectance changes due to the adsorption of vacuum residual gas on a cold mirror surface, which is the so-called cryo-pumping effect. This has only been experimentally estimated on a small-scale cryogenic system \cite{Miyoki2001}. A number of adsorbed molecules make an adlayer on a top of the cold mirror coating layers and cause changes in the mirror coating properties, such as reflectance, transmittance, and absorption depending on its properties and thickness. These effects can decrease the sensitivity of GWDs depending on the thickness of the adlayer. In order to characterize the adsorption process and to estimate the effects of adlayers on the KAGRA, we theoretically and numerically predicted the adlayer formation speed and its effects on a cooled mirror. Furthermore, we directly measured the formation of an adlayer on a cooled mirror using an optical cavity and discussed the possible reduction methods and the impacts on GWDs.\\
\quad This paper is structured as follows. First, we introduce the adlayer formation on a cooled mirror theoretically and then we show the result of the simulation of the adlayer formation on the KAGRA cryogenic mirrors in Sec.\ref{simulation}. Then, the effects of the adlayer as a mirror coating are theoretically discussed based on the thin film theory in Sec.\ref{sec.coating}. The details of the experiment are summarized in Sec.\ref{experiment} and the results are given in Sec.\ref{sec:result}. Finally, in Sec.{\ref{discussion}}, we discuss the method to decrease the molecular adlayer formation on cooled mirrors and the impacts of the adlayers on GWDs.

%%%%%%%%%%%%%%%%%%%%%%%%%%%%%%%%%%%%%%%%%%%%%%%%%%%%%%%%%%%
\section{Theory and Simulation}
\subsection{Molecules transfer and adlayer formation rate simulation}\label{simulation}
\begin{figure}
\centering
\includegraphics[width=8.0cm,clip]{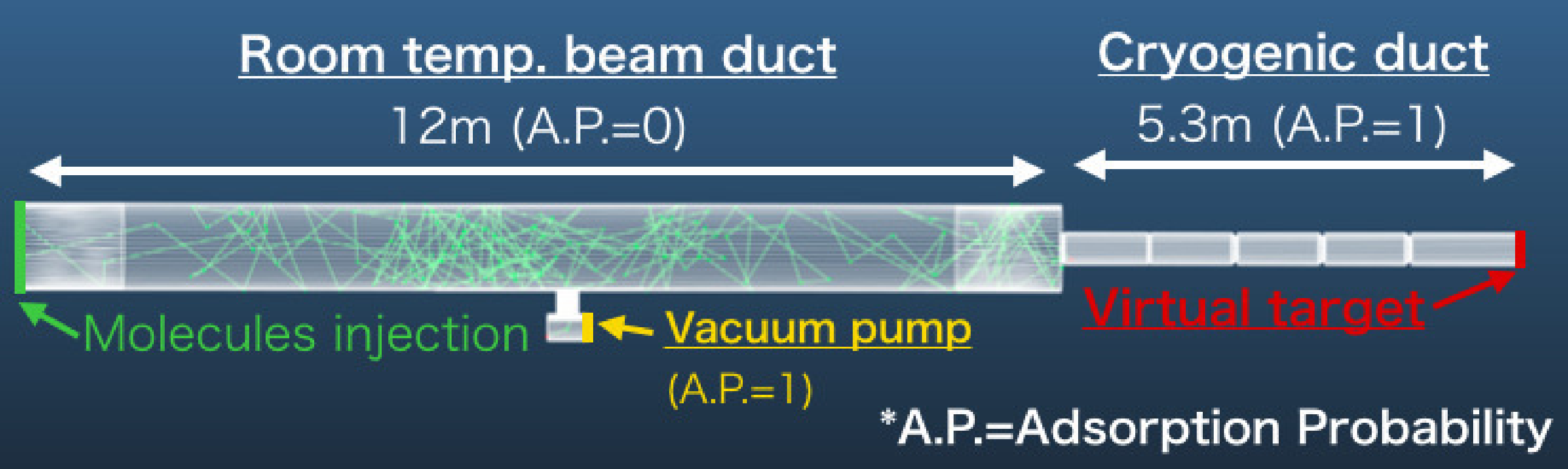}
\caption{Schematic view and structural dimensions for the MC simulation. The particles were injected from the left edge surface of the room-temperature duct and the particles that hit the virtual target were counted. The adsorption probability at the room-temperature beam duct was set to be zero, except for the vacuum pump parts, and set to be one for the cryogenic duct.}\label{fig:simu}
\end{figure}

First, let us consider the molecular accumulation rate on a cryogenic mirror surface. The molecular volumetric incident flux density $J$
 in a vacuum chamber with vacuum pressure $P$ is given by kinetic theory as :
  \begin{eqnarray}
  J = \frac{P}{\sqrt{2\pi m k_\mathrm{B} T}} k_{\mathrm{B}} T  = P \sqrt{ \frac{k_\mathrm{B}T}{2\pi m}}  \quad  \mathrm{ Pa \ m / s}, 
  \end{eqnarray}
  where $m$ is the mass of a molecule, $k_B$ is the Boltzmann constant, and $T$ is the temperature of the molecules. The number of molecules in a cryostat is less than that in a room-temperature vacuum duct, because the cryostat works as a type of vacuum pump. The molecules in the vacuum duct move to the cryostat to compensate the vacuum pressure difference. When molecules are transported from the long vacuum duct, the volumetric flow rate $Q$  is defined as $Q=C \Delta P$, where $C$ and $\Delta P$ are the conductance and pressure difference of the vacuum duct, respectively.
If the pressure inside the vacuum duct is sufficiently low and the molecules do not interact with each other, the conductance of a circular straight duct is written as \cite{Marquardt1999}:
 \begin{equation}\label{eq:conductance}
C=\frac{2\pi r^3}{3L} \sqrt{\frac{8 k_B T}{\pi m}}    \quad  \mathrm{m^3/s},
 \end{equation}
 where $r$ and $L$ are the radius and length of the duct, respectively. Then, the total volumetric molecular injection rate $Q_{\mathrm{tot}}$ is described as :
  \begin{eqnarray}\label{q_tot}
 \centering
 Q_{\mathrm{tot}} &=& JS +  C \Delta P  \nonumber \\
 &=& \sqrt{\frac{ k_B T}{2\pi m}}PS + \frac{2\pi r^3}{3L} \sqrt{\frac{8 k_B T}{\pi m}} \Delta P   \quad  \mathrm{Pa} \ \mathrm{m^3/s},
\end{eqnarray} 
where $S$ is the area of the cryogenic surface. The pressure level inside the cryostat is quite low, and thus the first term of Eq.(\ref{q_tot}) is neglectable. \\
\quad In the case of real vacuum ducts such as KAGRA or other GWDs, it is difficult to express the conductance of the vacuum system with a simple equation such as Eq.(\ref{eq:conductance}). 
To estimate the conductance of such a complicated duct, a Monte Carlo (MC) simulation works well \cite{Lobo2004}. The conductance can also be described as $C=C_0 \Gamma$, where $\Gamma$ is the transmission probability of the duct and $C_0= r^2  \sqrt{\pi k_{\mathrm{B}}T/2 m}$ is the opening conductance. In the MC simulation, the transmission probability $\mathrm{\Gamma}$, which is the ratio of particles between the input and output, can be numerically obtained.\\
\quad The adlayer formation rate can be defined by the number of adsorbed molecules on a cryogenic mirror per unit time. The probability of molecule adsorption, the so-called sticking coefficient, on cold surfaces for water molecules is almost unity below the freezing point \cite{Labello2011}. Then, the growth rate of the molecular adlayer $\eta$ is: 
 \begin{equation}\label{eq:2}
 \eta = \mathrm{K} \frac{m Q_\mathrm{tot}}{S \rho k_{\mathrm{B}}T}   \sim \mathrm{K} \frac{m C_0 \mathrm{\Gamma} \Delta p}{S \rho k_{\mathrm{B}}T}   \quad  \mathrm{m/s},
 \end{equation}
 where K is the geometrical factor, which is the probability of molecules hitting the mirror surface through the cryogenic duct and $\rho$ is the density of the molecules. Usually, the percentage of the water molecules in vacuum chambers is large and the water molecules slowly accumulate on the cryogenic mirror surface. In such a situation, the water molecules form a low-density amorphous state (LDA), which has a density of $\rho = 0.82$ $\mathrm{g/cm^3}$ \cite{Westleya1998}. \\
 \quad In order to evaluate the molecular accumulation rate by the MC simulation, Molflow+, which is a MC simulator package developed at CERN \cite{Molflow}, was used. Figure \ref{fig:simu} shows the schematic view and the structural dimensions for the MC simulation. In this simulation, the particles were injected from the left edge surface of the room-temperature duct illustrated in Fig.\ref{fig:simu}, and the particles that hit the vacuum pump and the cryogenic duct surface were assumed to be perfectly adsorbed. Then, the number of particles that come through the cryogenic duct were counted. The geometrical factor K was also obtained with this simulation. The values of each parameter obtained from this MC simulation, assuming the KAGRA vacuum duct, are summarized in Table \ref{MC}. Accordingly, the adlayer growth rate $\eta$ was calculated to be $42$ nm/day for the LDA state of the water molecules.

  \begin{table}[t]
  \caption{Results of Monte Carlo simulation \label{MC}}
\begin{tabular}{ l | c | c | c}
\hline \hline
Name & Symbol &  &Unit  \\
\hline
Geometrical factor  & K & 0.014 & - \\
Transmission probability  & $\mathrm{\Gamma}$ & $ 0.00139 $ & - \\
Opening conductance & $C_0$ & 116.6 & $\mathrm{m^3/s}$ \\
Conductance  & C & 0.162 &$\mathrm{m^3/s}$ \\
Pressure difference & $\Delta P$ & $6.9 \times 10^{-6}$ & Pa \\
\hline
Layer formation rate & $\eta$ & 42 & $\mathrm{nm/day}$ \\
 \hline \hline
\end{tabular}
\end{table}

 %%%%%%%%%%%%%%%%%%%%%%%%%%%%%%%%%%%%%%%%%%%%%%%%%%%%%%%%%%%
 \subsection{Molecular adlayer as an optical coating}\label{sec.coating}
\begin{figure}
\centering
\includegraphics[width=8.0cm,clip]{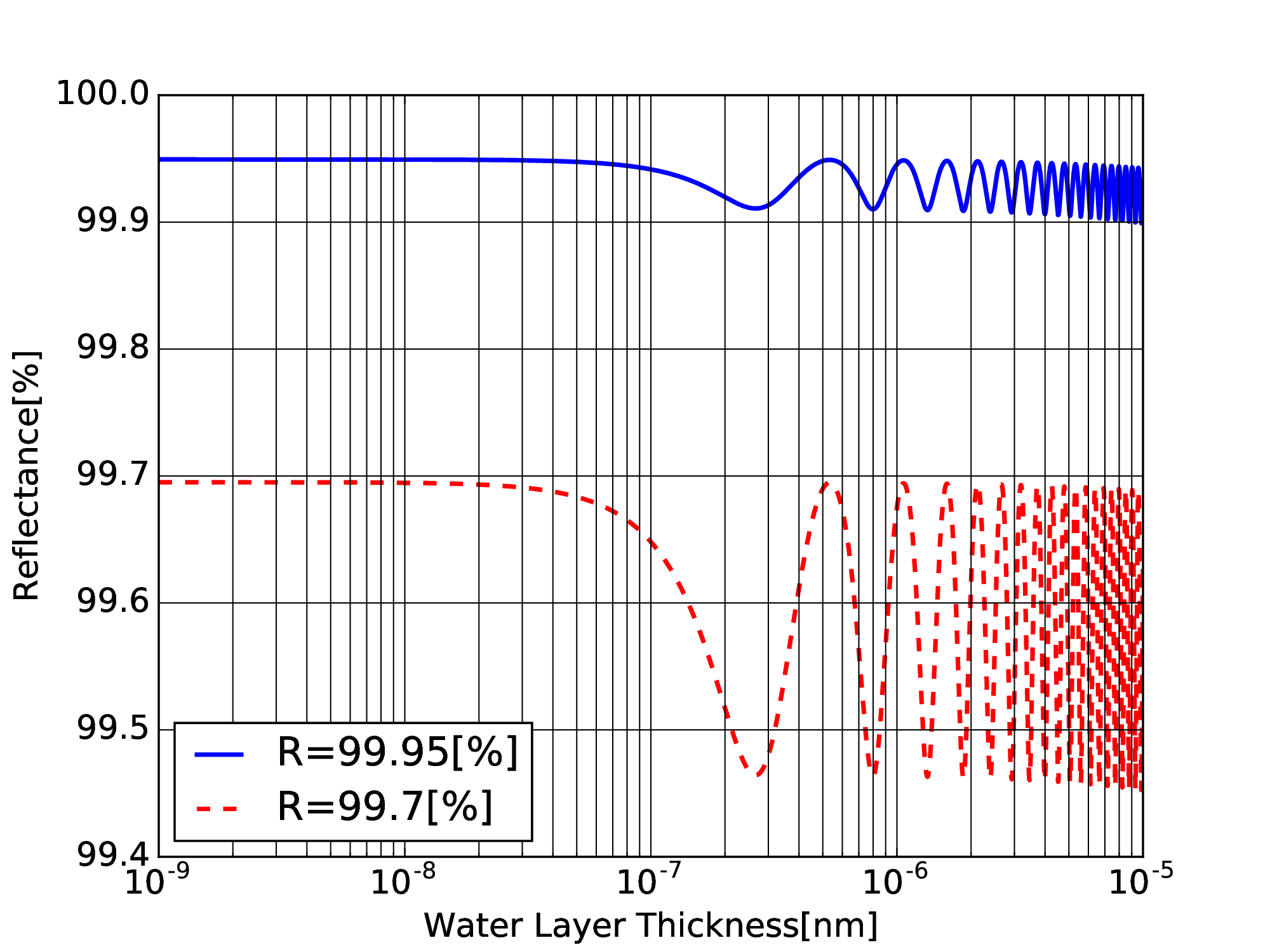}
\caption{Reflectance changes induced by molecular adlayer growth. Depending on the adlayer thickness, the reflected light from the top of molecular layer and multilayer coating repeat constructive and destructive interference. As a result, the reflectance of the mirror oscillates. The amplitude of this oscillation is maximized when the amplitudes of the two reflected lights are same, and the absorption in the molecular adlayer decreases the reflectance gradually. The blue line and red dashed line show the calculation results with coatings that originally have power reflectances of $99.95 \%$ and $99.7 \%$, respectively. }\label{fig:1}
\end{figure}

 In GWDs, various mirrors that have different reflectances are used according to their roles. A multilayer coating technique is used to control the reflectivities of these mirrors. A multilayer coating film is composed of low- and high-refractive-index materials that are stacked alternately, and the refractive index and thickness of each layer determine the reflectance for specific laser wavelengths \cite{John1990}.
  To calculate the reflectance or transmittance of a multilayer coating film, a characteristic matrix is used. In the case of materials with a complex refractive index $N_\mathrm{i}$ and mechanical thickness $d_\mathrm{i}$, the characteristic matrix is described as:
 \begin{equation}
 M_i=
 \begin{bmatrix}
\cos{\delta_\mathrm{i}}  &  i \sin{\delta_\mathrm{i}}/N_\mathrm{i}\\
iN_\mathrm{i}\sin{\delta_\mathrm{i}} & \cos{\delta_\mathrm{i}}
 \end{bmatrix},
 \end{equation}
 where $\delta_\mathrm{i}=2\pi N_\mathrm{i}d_\mathrm{i}/\lambda$ is an optical phase shift induced in one layer. The total matrix $M_{\mathrm{tot}}$ and its elements $m_{\mathrm{ij}}$ are defined as:
\begin{align}\label{eq:multi-layer}
 M_{\mathrm{tot}}&=M_{\mathrm{1}} \cdot M_{\mathrm{2}} \cdots M_{\mathrm{n-1}} \cdot M_{\mathrm{n}} \nonumber \\
&= \prod_{\mathrm{k}=1}^{\mathrm{n}}M_{\mathrm{k}}=
 \begin{bmatrix}
 m_{11} &im_{12}\\
 im_{21} & m_{22}
%a_{11}+ib_{11}  & a_{12}+ib_{12} \\
%a_{21}+ib_{21} & a_{22}+ib_{22}
 \end{bmatrix}.
 \end{align}
The characteristic matrix of a multilayer coating film is given as:
\begin{align}\label{eq:4}
 \begin{bmatrix}
 B \\ C
 \end{bmatrix}
=
 \begin{bmatrix}
 m_{11} &im_{12}\\
 im_{21} & m_{22}
 \end{bmatrix}
 \begin{bmatrix}
 1 \\ n_m
 \end{bmatrix},
 \end{align}
 where $B$ and $C$ are elements of the characteristic matrix, and $n_\mathrm{m}$ is the refractive index of the substrate material. Eq.(\ref{eq:4}) gives the Fresnel coefficient of reflected light as $\rho = \left (n_{0} B - C \right ) / \left ( n_{0} B +C\right )$. The Fresnel coefficient directly leads to the power reflectance of light as $R= \left | \rho \right | ^2$.\\
\quad Accumulated molecules form an adlayer, which has refractive index value. This means that the adlayer works as an optical coating. In order to include its effect, one can add the characteristic matrix of the molecular adlayer in Eq.(\ref{eq:multi-layer}). If there is an adlayer on the top of the multilayer coating film, the reflected light from the top of the growing adlayer and the existing multilayer coating repeat the constructive and destructive interference according to the adlayer thickness. This optical interference results in an oscillation of the reflectance. In addition to this, the reflectance inevitably decreases because of the optical absorption inside the molecular adlayer. Figure \ref{fig:1} shows an example of the reflectance with a growing adlayer of water molecules. The total reflectance of the mirror oscillates owing to continuous change in the interference condition depending on the adlayer thickness. The amplitude of the reflectance oscillation is approximately determined by the ratio between the reflected light power from the top of the adlayer and the multilayer coating film. Therefore, the amplitude is maximized when the amplitude of two reflected lights are almost equal. Because the amplitude reflectivity of the water adlayer is approximately $10 \%$, a multilayer coating film that has an amplitude reflectance of approximately $10 \%$ is significantly affected by the molecular adlayer.
 
%%%%%%%%%%%%%%%%%%%%%%%%%%%%%%%%%%%%%%%%%%%%%%%%%%%%%%%%%%%
\section{Experimental method}\label{experiment}
 In order to estimate the reflectance change due to the adlayer on a cryogenic mirror, a FP cavity was installed inside the KAGRA cryostat and cooled to approximately 47 K. The sharpness of the FP cavity resonance is evaluated by its finesse $\mathcal{F}$:
\begin{equation}
\mathcal{F} =  \frac{\pi \sqrt{r_1 r_2}}{1-r_1r_2} ,
\end{equation} 
where $r_1$ and $r_2$ are the amplitude reflectances of the FP cavity mirrors. As explained in Sec.(\ref{sec.coating}), the molecular adlayer changes the mirror reflectance, transmittance, and absorption of light according to its complex refractive index, density, and thickness. Therefore, by monitoring the change in finesse, these material properties can be estimated and the component of the adlayer can be identified.\\
\quad There are several methods to measure the finesse. For example, the ringdown method \cite{Paul2001}, cavity transfer function method \cite{Uehara1995a,Uehara1995}, and doppler effect method \cite{Isogai2013a}. In this experiment, the ringdown method was used to reduce the measurement time and to avoid heating up of the mirror due to absorption.

%%%%%%%%%%%%%%%%%%%%%%%%%%%%%%%%%%%%%%%%%%%%%%%%%%%%%%%%%%%
\subsection{Ringdown method for finesse measurement}
\begin{figure}[htb]
\centering
\includegraphics[width=8.0cm,clip]{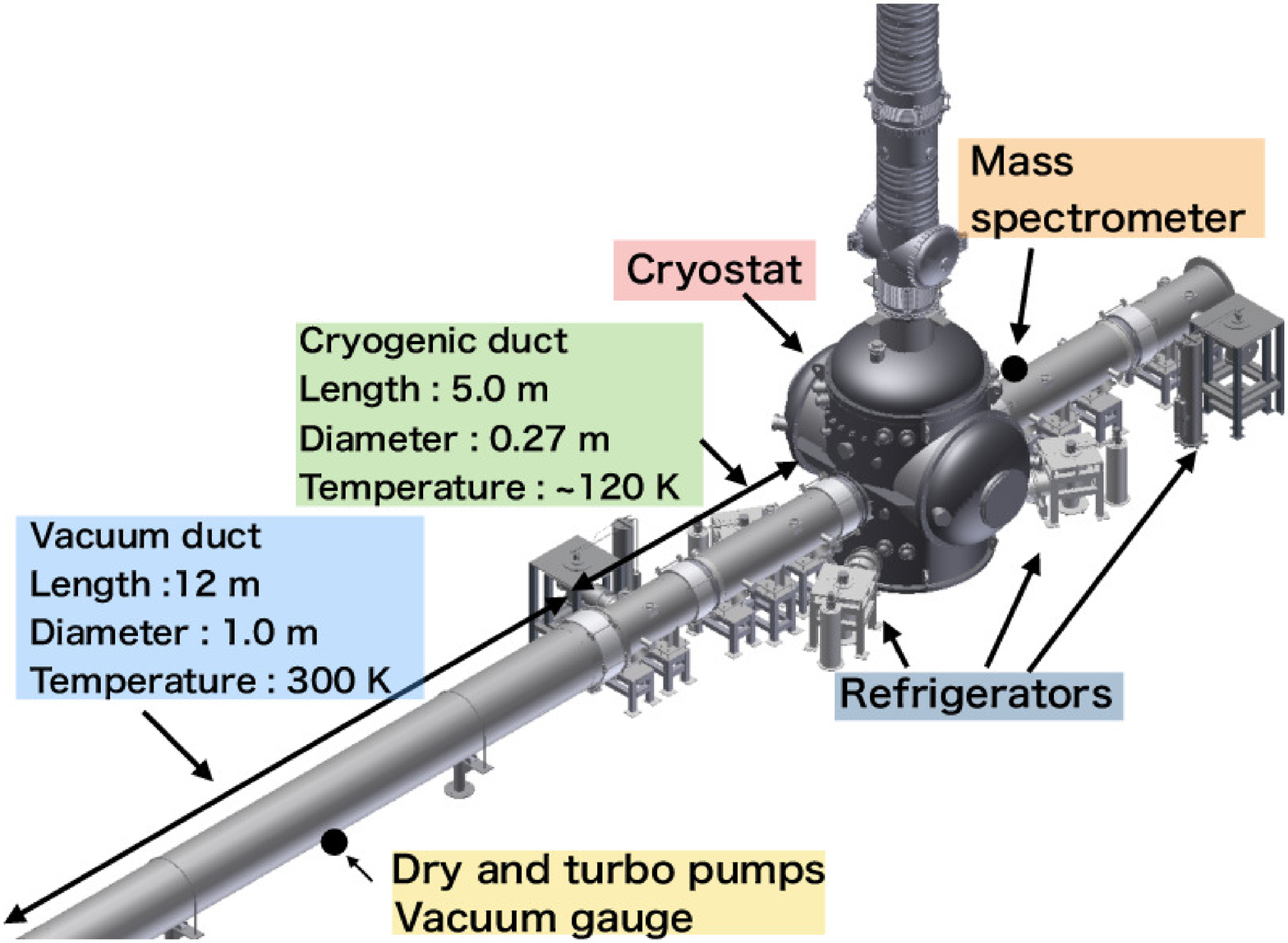}
\caption{Drawing of setup around the KAGRA cryostat. During the experiment, the vacuum duct is separated by gate valves, which are placed at 17 m from the cryostat. The lengths of the vacuum and cryogenic ducts are 12 m and 5 m, respectively. There are six pulse-tube refrigerators for a KAGRA cryostat. Four 4-K refrigerators are used to cool mirror and cryostat, and two 80-K refrigerators are used to cool two 5-m cryogenic ducts \cite{Tokoku2014}. A set of dry and turbo-molecular vacuum pumps is attached at the middle of vacuum duct on each side and a vacuum gauge was also attached. To identify the type of residual molecules, a mass spectrometer was attached near the cryostat.}\label{fig:vac}
\end{figure}

\begin{figure}[t]
\centering
\includegraphics[width=8.0cm,clip]{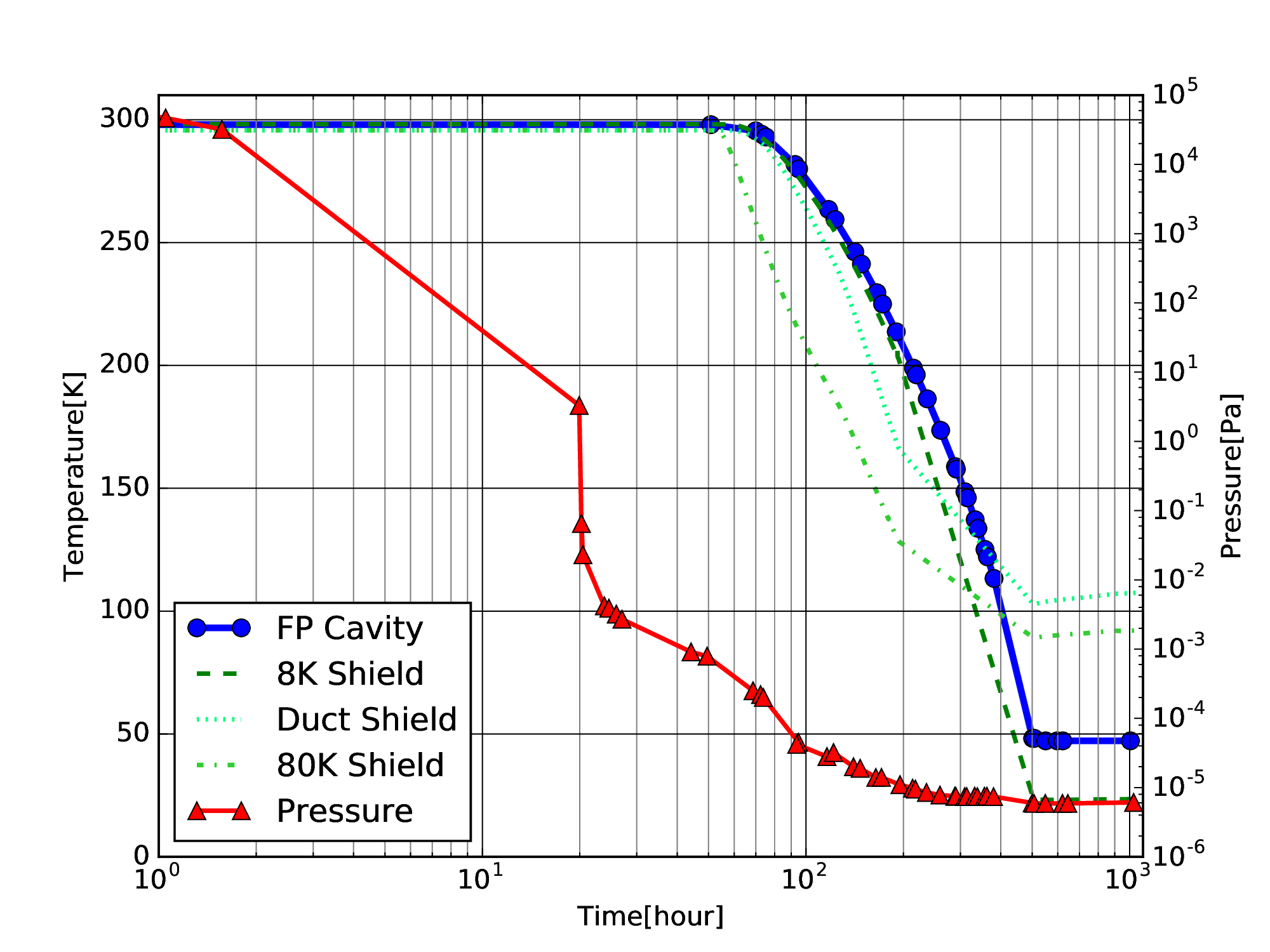}
\caption{Vacuum pressure and temperature change around the FP cavity. The blue line and three green dashed lines show the temperature of FP cavity, bottom of 8-K radiation shield, cryogenic duct, and bottom of the 80-K radiation shield, respectively. The red line shows the pressure measured at the middle of the room-temperature beam duct. The position of the vacuum gauge is shown in Fig.\ref{fig:vac}. Finally, the vacuum pressure reached $6.0 \times 10^{-6}$ Pa after the cooling process was complete.}\label{fig:2}
\end{figure}

By injecting intensity-modulated laser light with rectangular waves, the transient response of the FP cavity can be observed in the transmitted light from the FP cavity. The time constant (or the optical storage time) of the FP cavity can be obtained from this transient response curve. This method is called the ringdown method. The relation between the time constant and the finesse $\mathcal{F}$ of a FP cavity is described as:
\begin{equation}\label{eq:finesse}
 \tau_\mathrm{s}=\frac{2L}{\pi c} \mathcal{F} ,
\end{equation}
where $\tau_\mathrm{s}$, $c$, and $L$ are the time constant, speed of light, and length of the FP cavity, respectively. 
The transient response curve depends on the modulation depth $\beta$ of the injected laser power. The incident laser beam field $E_{\mathrm{i}}(t)$ with a modulation of the rectangular waves at $t=0$ can be written as: 
\begin{equation}
E_\mathrm{i}(t)=
\left\{
\begin{array}{l}
|E_\mathrm{i}|e^{i\omega t}   \qquad \qquad t< 0\\
|E_\mathrm{i} - \beta|e^{i\omega t}   \quad \quad 0 \le t \quad , \\ 
\end{array}%
\right.
\end{equation}
where $\omega$ is the angular frequency of light. The transmitted laser beam field $E_{\mathrm{t}}(t)$ is calculated by considering the transient response:
\begin{equation}
E_{\mathrm{t}}(t) = E_{\mathrm{t}}(0)-H_{\mathrm{FP}} \beta \left ( 1-e^{-\frac{t}{\tau_{\mathrm{s}}}}      \right ) ,
\end{equation}
where $H_{\mathrm{FP}}$ is the ratio of the injected and transmitted laser power to the FP cavity. Hence, the power of the transmitted light can be written as:
\begin{equation}
P_{\mathrm{t}}(t) =
\left\{
\begin{array}{l}
 |H_\mathrm{FP}|^2  \left ( \left |E_\mathrm{i} \right |^2  -2 \left |E_\mathrm{i} \right | \beta \left(1-e^{-\frac{t}{\tau_\mathrm{s}}} \right ) \right)    \qquad (\beta \ll 1)   \\
|H_\mathrm{FP}|^2  \left |E_\mathrm{i} \right |^2 e^{-\frac{2t}{\tau_\mathrm{s}}} \hspace{30mm}   (\beta \sim E_\mathrm{i}) ,
\end{array}%
\right.
\end{equation}
Consequently, we can obtain the time constant from the transient response for any modulation depth. In this experiment, the incident laser beam was completely shut down by an acoustic optic modulator (AOM). Hence, the modulation depth was the same as the amplitude of the injected laser electric field and the detectable time constant was $\tau_{\mathrm{s}}/2$.

 %%%%%%%%%%%%%%%%%%%%%%%%%%%%%%%%%%%%%%%%%%%%%%%%%%%%%%%%%%%
subsection{KAGRA cryostat and vacuum systems}
 KAGRA has four cryostats to cool sapphire test masses inside. In this experiment, one of the cryostats was used. 
Figure \ref{fig:vac} shows an image of the setup around the KAGRA cryostat with six cryocoolers, two radiation shield ducts extended from the cryostat in opposite directions and vacuum ducts connected with these radiation shield ducts. The lengths of the vacuum and cryogenic ducts are 12 m and 5 m, respectively. A KAGRA cryostat has four pulse-tube refrigerators: two cryocoolers cool a sapphire mirror and its suspension system and the other two cryocoolers cool two-layer thermal radiation shields. The 5 m of radiation shield ducts, which are called the cryogenic ducts, have two 80-K refrigerators \cite{Tokoku2014}. A set of dry and turbo-molecular vacuum pumps was attached at the middle of the vacuum duct on each side and a vacuum gauge was also attached. The vacuum pressure reached $7.9\times10^{-4}$ Pa before starting cool. After a sufficient pumping duration, the entire cryogenic systems were cooled down, and the vacuum pressure reached $6.0\times 10^{-6}$Pa at the middle of vacuum duct. Figure \ref{fig:2} shows the vacuum pressure and the temperature change around the FP cavity that was set inside the cryostat. \\
\quad During the finesse measurement, the components of residual molecules were also monitored by a mass spectrometer attached near the cryostat. The main detected molecules were $\mathrm{H_2O}$, O, and OH, where O and OH can be generated from $\mathrm{H_2O}$. Because the partial pressure of $\mathrm{H_2O}$ molecules at 50 K is so low, it is easily predictable that most of the $\mathrm{H_2O}$ molecules were adsorbed on cryogenic parts including mirror surfaces of the FP cavity.

%%%%%%%%%%%%%%%%%%%%%%%%%%%%%%%%%%%%%%%%%%%%%%%%%%%%%%%%%%%
 \subsection{Fabry--Perot cavity and optics}
\begin{figure}[htb]
\centering
\includegraphics[width=8.0cm,clip]{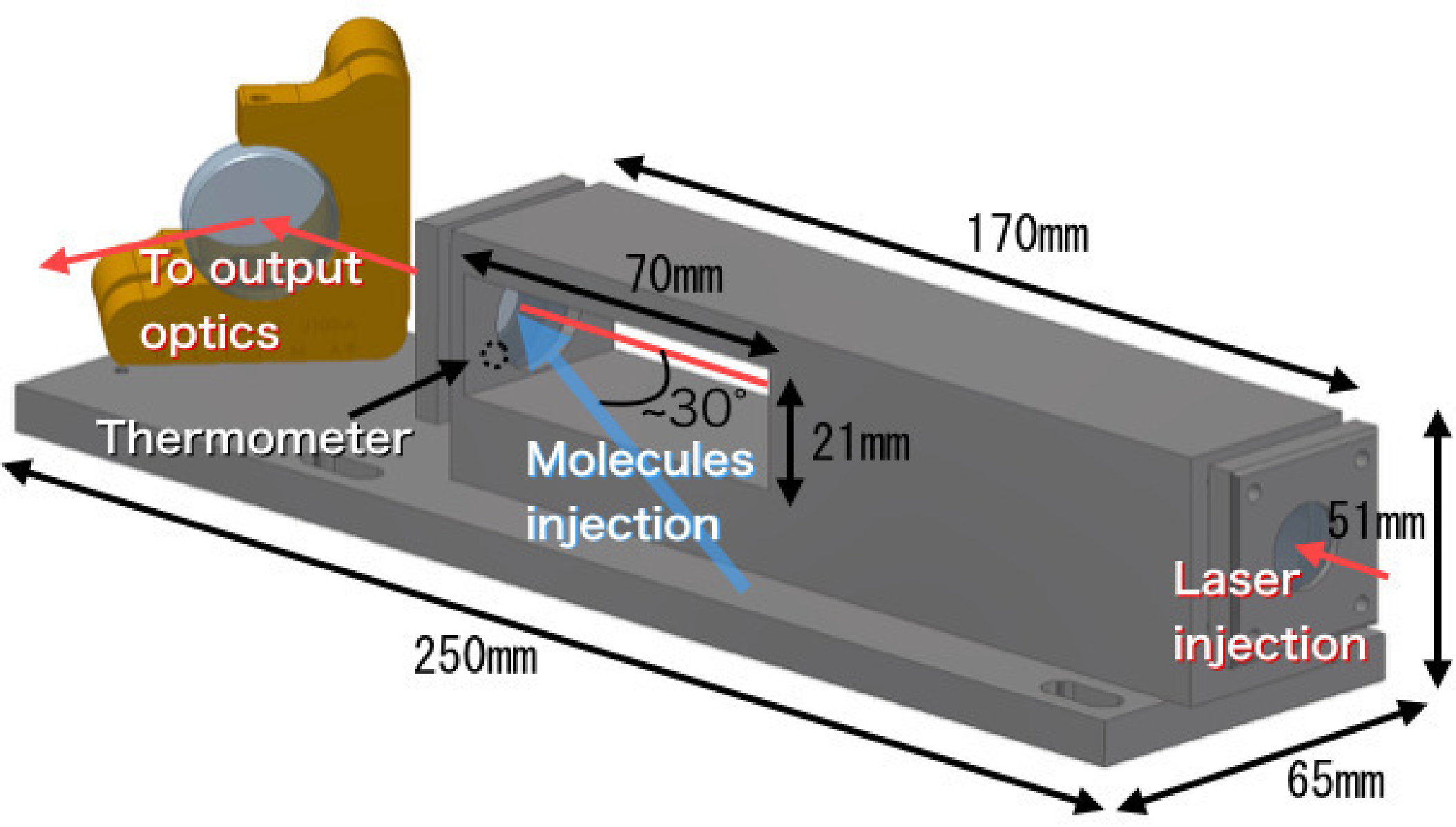}
\caption{Design of the FP cavity. The spacer is made of 316L stainless steel and the cavity length is 17 cm. This spacer has a side window to expose the target cavity mirror to the room-temperature beam duct, from where molecules come. The other side of the mirror did not face the room-temperature area, and thus the molecular adlayer is formed only on the former mirror and the change of finesse is induced by the adlayer on the target mirror. The temperature of the FP cavity was monitored by a thermometer that was attached near the target mirror.}\label{fig:spacer}
\end{figure}

\begin{figure}[htb]
\centering
\includegraphics[width=8.0cm,clip]{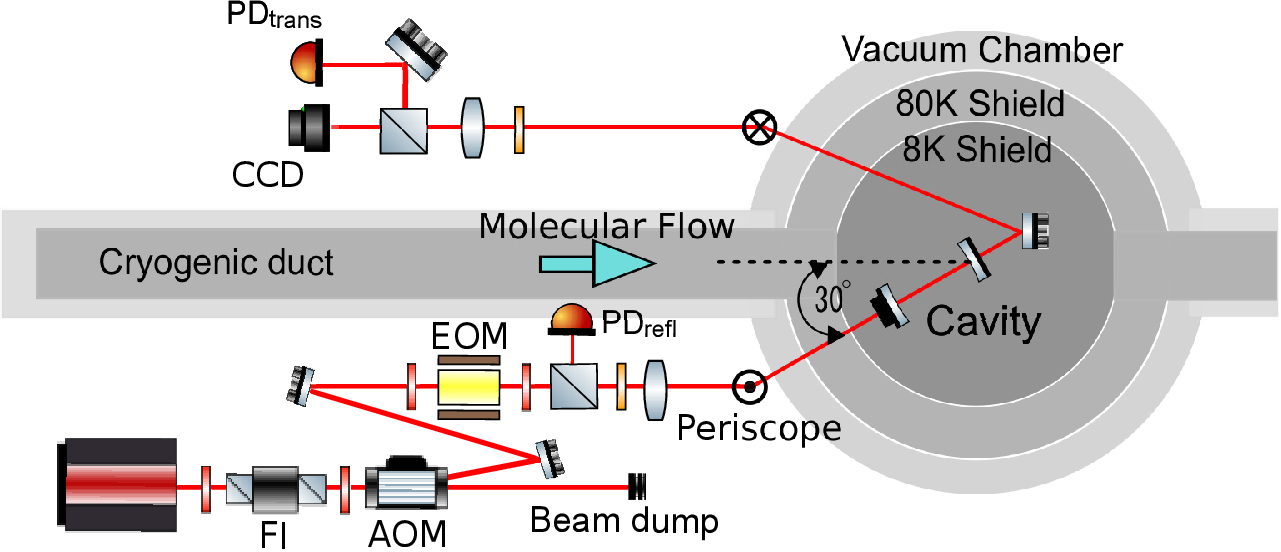}
\caption{Experimental setup. Input optics were constructed to control the laser frequency to maintain the resonance of the FP cavity and to shut down the injected laser beam rapidly by the AOM. The injection angle of the laser to the FP cavity is approximately $30 ^\circ$ tilted from the beam axis of KAGRA. FI: Faraday isolator, AOM: acoustic optic modulator, EOM: electro optic modulator, $\mathrm{PD}_{\mathrm{refl}}$: photodetector to detect the reflected light from the FP cavity, $\mathrm{PD}_{\mathrm{trans}}$: photodetector to detect thfe transmitted light from the FP cavity.}\label{fig:setup}
\end{figure}

Figure \ref{fig:spacer} shows the design of the FP cavity. The spacer is made of 316L stainless steel and the cavity length is 17 cm. 
The spacer has side windows to expose a cavity mirror to the beam duct from where molecules come. The other side of the mirror did not face the room-temperature area, and thus the molecular adlayer is formed only on the former mirror. The temperature of the FP cavity was monitored by a thermometer attached near the target mirror in Fig.\ref{fig:spacer}.\\
\quad Figure \ref{fig:setup} shows the experimental setup of the finesse measurement of the FP cavity. In order to adjust the height of the FP cavity mirrors to that of the KAGRA sapphire mirrors, the FP cavity was set on a stainless-steel table, and connected to the bottom of the inner radiation shield of the cryostat by pure aluminum heat links to maintain sufficient thermal conductivity. The Pound--Drever--Hall (PDH) method was utilized to keep the laser light resonating inside the FP cavity. The error signal was fed back to the laser frequency tuning PZT. 
To shut down the input laser power quickly and completely, an AOM was introduced and the first-order laser beam was injected to the FP cavity.
The time constant ($\tau_s$) of the FP cavity was measured by monitoring the transmitted light, which includes the transient response of the FP cavity, with a high-speed photodetector.
It was confirmed that the shutter speed of the AOM and the response of the photodetector were lower than the expected time constant of the FP cavity. The time constant of the FP cavity was measured to be $\tau_\mathrm{s} =17.7 \pm 0.75 $ $\mu \mathrm{s}$ and the resulting finesse was $49000 \pm 2100$ before cooling.\\
 
 %%%%%%%%%%%%%%%%%%%%%%%%%%%%%%%%%%%%%%%%%%%%%%%%%%%%%%%%%%%
 \section{Results}\label{sec:result}
\begin{figure}[htb]
\centering
\includegraphics[width=8.0cm,clip]{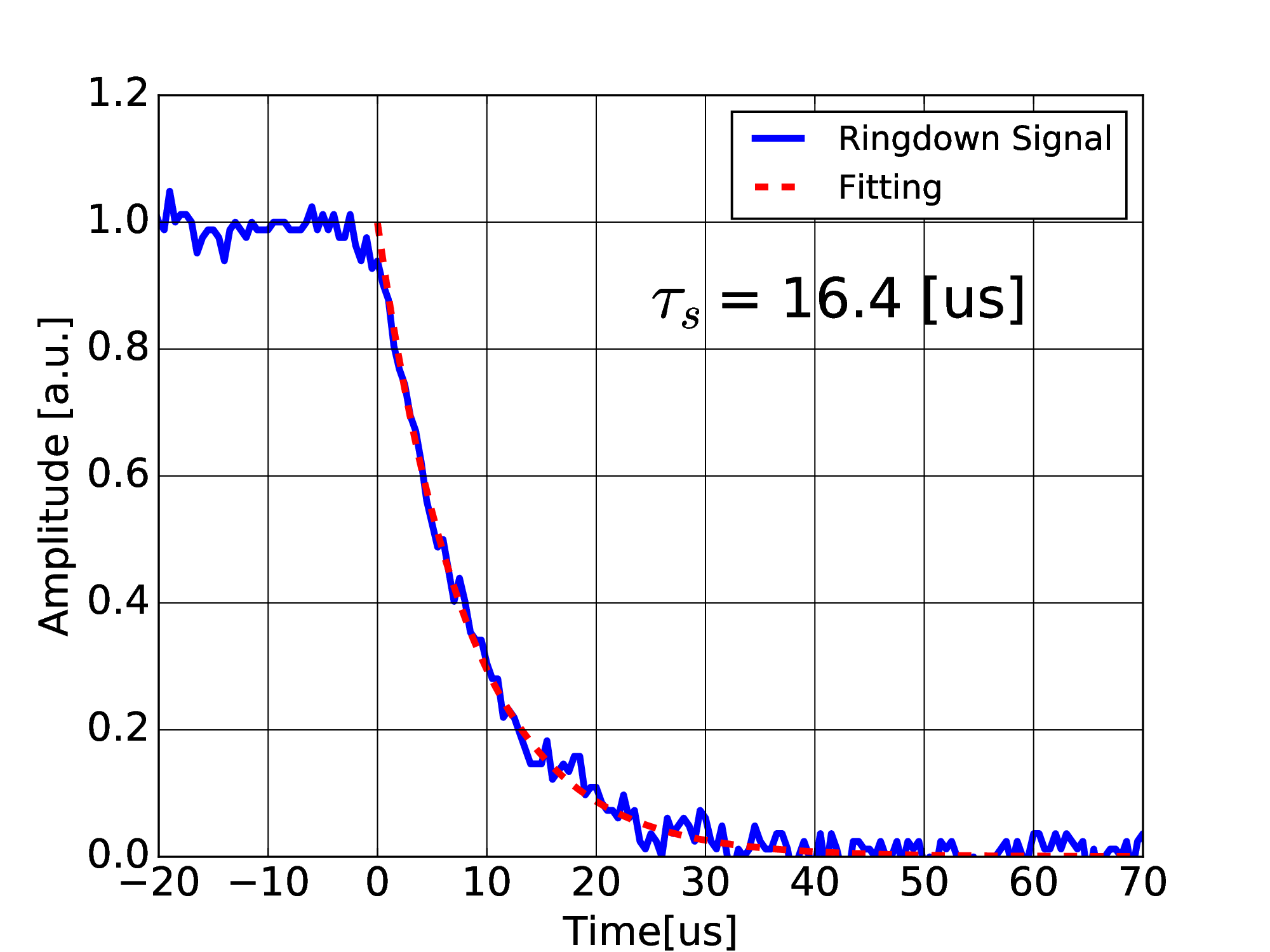}
\caption{Sample of the ringdown signal. The injected laser power was turned off at 0 s by the AOM and the stored laser power started to decrease. The blue line and the red dashed line show the measured ringdown signal and the fitting curve, respectively. In this case, the time constant was 8.2 $\mu$s, which means that the FP cavity had a time constant of $\tau_\mathrm{s}=16.4$ $\mu$s.}\label{ringdown}
\end{figure}

\begin{figure}[htb]
\centering
\includegraphics[width=8.0cm,clip]{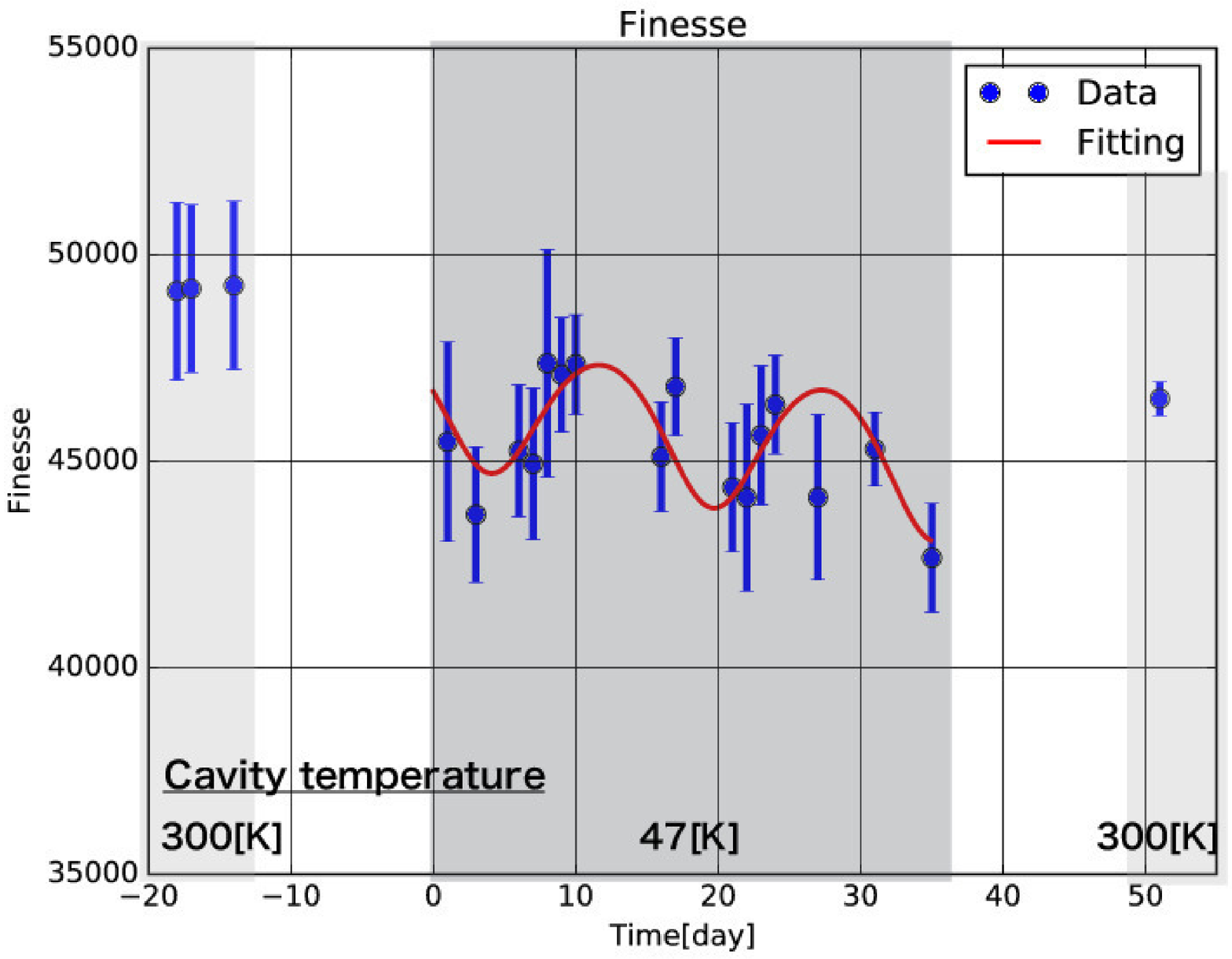}
\caption{Estimated finesses. The light gray and the dark gray areas show the periods when the temperature of the FP cavity was 300 K and 47 K, respectively. The blue circles are the averaged finesses and the error bars indicate the statistical error. To calculate the average finesse and statistical error, 20 times of the ringdown measurements are used. The red curve is the fitting line. In order to calculate the original coating of a mirror, the materials with low and high refractive index were stacked on a fused silica substrate alternately and the parameters of $\mathrm{SiO_2}$ and $\mathrm{Ta_2O_5}$ were used for the materials. Each layer has a quarter wavelength of optical thickness and the total number of the quarter-wavelength layers is 38. In this fitting function, the molecular adlayer was stacked on this coating and the adlayer formation speed, initial phase, initial optical loss, and complex refractive index were the fitting parameters.}\label{result}
\end{figure}

Figure \ref{ringdown} shows a typical ringdown signal from the FP cavity. The blue line and the red dashed line show the measured data and the fitting curve, respectively. The ringdown signals were taken by a photodetector to monitor the transmitted laser power from the FP cavity and the time constant was obtained by fitting the ringdown data. The finesse values derived from the measured ringdown signals of the FP cavity are shown in Fig.\ref{result}. The finesse was calculated according to Eq.(\ref{eq:finesse}), assuming $L=17~\mathrm{cm}$. In Fig.\ref{result}, the blue circles show the average finesse during a day and the error bars correspond to the statistical errors. To calculate the average finesse and statistical error, 20 times of the ringdown measurements are used. The red line shows the fitting line and the fitting results are summarized in Table \ref{fitting}. \\
\quad The difference in finesse before and just after cooling is expected to be due to the initial huge amount of molecular adsorption or the FP cavity axis change induced by the thermal shrink of the rigid spacer. These two effects cannot be distinguished by this measurement.
After reaching a stable cryogenic temperature, the temperature fluctuation of the FP cavity was 0.03 K. Therefore, the cavity length fluctuation induced by the temperature fluctuation was too small to explain the measured reflectance oscillation. As shown in Table \ref{fitting}, the calculated sticking rate of molecules to the mirror is 27 $\pm$ 1.9 $\mathrm{nm/day}$. The real and imaginary part of refractive index of the adlayer was $1.26 \pm 0.073$ and $2.2 \times 10^{-7} \pm 1.3 \times 10^{-7} $. The real and imaginary part of the refractive index of the Ih state of water for 266 K are reported as 1.32 and $1.22 \times 10^{-6} $ \cite{Warren2008}. The real part of the measured refractive index is comparable with the reported value, but there are about 5 times of difference in the imaginary part of the refractive index. The refractive index of the LDA state of the water could be changed depending on its formation process \cite{Labello2011}, and the study of the complex refractive index for the LDA water have never completed for 1064 nm of the wavelength. In order to determine the refractive index of the LDA water for 1064 nm of wavelength more precisely, a further experiment is needed.

%%%%%%%%%%%%%%%%%%%%%%%%%%%%%%%%%%%%%%%%%%%%%%%%%%%%%%%%%%%
\section{Discussions}\label{discussion}
The simulation discussed in the Sec.{\ref{simulation}} contains some uncertainties. In order to improve the accuracy of the simulation, we need to investigate the aspects below: \\
\quad (1) Vacuum pressure simulation. In order to estimate the molecular adlayer growth rate, the vacuum pressure distribution was also simulated by using an experimental value measured at the middle of the vacuum duct. In this simulation, it was assumed that the outgassing from the vacuum duct surfaces was homogeneous. The actual pressure distribution is expected to be different from this calculation, owing to inhomogeneous outgassing from the vacuum chamber surfaces and super insulators. Additionally, the inhomogeneous and unexpected temperature distribution of the cryogenic duct could possibly change the vacuum pressure profile. These uncertainties can vary the adlayer growth rate.
In order to measure the vacuum pressure inside vacuum chambers with better accuracy, a vacuum pressure measurement based on an optical cavity has been developed at NIST \cite{Hendricks2015}. By applying this technique to KAGRA, the vacuum pressure at each point of the beam duct and the cryostat can be measured. Finally, this measurement possibly leads to better characterization of the adlayer formation on cooled mirrors in KAGRA.\\
\quad (2) Molecular reflection at the cryogenic parts: Typically, the partial vapor pressure of water molecules at cryogenic surfaces is quite low, and therefore the reflection at that surface could be negligible. However, around 120 K, such as in the KAGRA cryogenic ducts, molecules can be released within a finite time depending on the surface potential depth. In order to evaluate this molecule release effect, the physisorption energy depth of the KAGRA cryogenic ducts surfaces should be measured, and the conductance of the cryogenic ducts including the molecular partial adsorption, or release in finite time, must be included in the simulation.\\
\quad In the above sections, we discussed the adlayer effects based on the thin film theory. The entire optical loss was considered as the result of absorption inside the adlayer. However, the scattering caused by the surface roughness can also be the source of an optical loss, and it can cause troublesome noise known as ``stray light noise" or ``scattering noise" in GWDs \cite{Vinet1996,Zeidler2017}. Further theoretical and experimental studies are needed to characterize the scattering effects caused by an adlayer.

\begin{table}[t]
\caption{Fitting results}\label{fitting}
\begin{tabular}{ l | l | l }
 \hline \hline
Sticking rate & $27.1 \pm 1.9$ & nm/day \\
Initial phase & $0.81 \pm 0.26 $ & rad \\
Initial optical loss & $ 17 \pm 1.8 $ & ppm \\
Refractive index (Real)& $1.26 \pm 0.073 $\\
Refractive index  (Imaginary)& $ 2.2 \times 10^{-7} \pm 1.26 \times 10^{-7} $ \\
\hline \hline
\end{tabular}
\label{tab:2}
\end{table}

%%%%%%%%%%%%%%%%%%%%%%%%%%%%%%%%%%%%%%%%%%%%%%%%%%%%%%%%%%%
\subsection{Possible reduction methods of molecular adlayer formation}
In order to maintain the sensitivity of KAGRA during its observation, we need to mitigate the accumulation of molecules and remove the accumulated molecules. Both passive and active desorption methods are conceivable. The passive reduction methods are as follows: (1) Better vacuum condition in the room-temperature vacuum duct: the molecular adlayer formation rate due to the molecular transportation through the room-temperature vacuum duct is proportional to the vacuum pressure difference between the cryostat and the vacuum duct, as shown by Eq.(\ref{eq:2}). In this experiment, the vacuum pressure at the vacuum duct was relatively high and it caused rapid adlayer formation. If it was a designed value ($2.0\times 10^{-7}$ Pa), we can expect the formation rate to decrease by approximately a factor of 50. (2) Longer cryogenic duct: The solid angle of $\Omega$ from the center of the mirror to the room-temperature region is calculated as:
\begin{equation}\label{eq:solidangle}
\Omega = 2\pi \left( 1- \frac{z}{\sqrt{z^2+r^2}}   \right) ,
\end{equation}
where $z$ is distance from the mirror to the edge of cryogenic duct and $r$ is the radius of the duct. In the present KAGRA case, the length of the cryogenic duct is 5 m and the diameter is 270 mm. By applying these values to Eq.(\ref{eq:solidangle}), we can find that our cryogenic duct have approximately 2 msr of the solid angle for the 300 K region. If we could use a cryogenic duct of length  30 m, the solid angle would decrease to 0.2 msr, which has the same effect as a ten-fold reduction in the vacuum pressure from the point of view of the injection rate of molecules to the mirror surface.\\
\quad To date, several active desorption methods have already been studied and developed from the perspectives of vacuum engineering, surface material physics, and chemistry \cite{Focsa2006,Fraser2001}. The photo-desorption method is one of these. By illuminating the mirror surfaces with a laser beam whose wavelength corresponds to the frequency of internal molecular vibrational motion, the internal mode can be excited and finally the molecules can get enough energy to escape from the surface potential. Among the internal vibrational modes of a $\mathrm{H_2O}$ molecule, symmetric and antisymmetric stretching vibrations have strong absorption. The corresponding wavelength is approximately 3 $\mathrm{\mu m} $ at the cryogenic temperature \cite{Focsa2006}. In order to remove the adsorbed molecules by this method, the laser power density should be high and threshold power density is estimated as 375 $\mathrm{kW/cm^2}$ \cite{Labello2011} for the 3.0 $\mathrm{\mu m}$ of the wavelength. However, the high-power-density laser would damage the coating. In order not to damage the coating, by heating up the mirror slightly and increase the energy of the molecules in advance, the threshold power density to remove the molecules could be decreased.

%%%%%%%%%%%%%%%%%%%%%%%%%%%%%%%%%%%%%%%%%%%%%%%%%%%%%%%%%%%
\subsection{Impacts of the adlayer on KAGRA}
\begin{figure}[htb]
\centering
\includegraphics[width=8.0cm,clip]{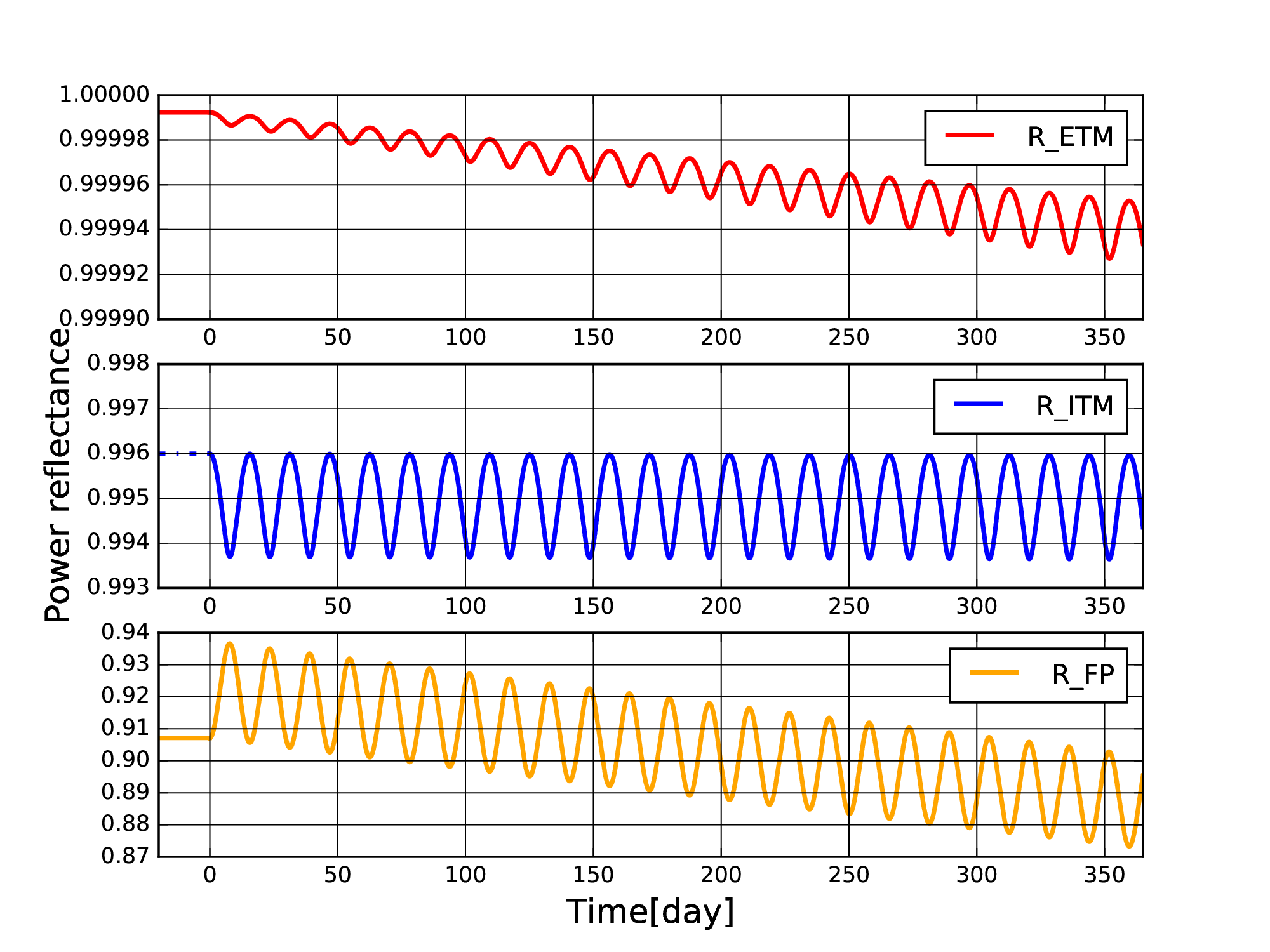}
\caption{Reflectance oscillations of ITM (top), ETM (middle), and the arm FP cavity (bottom) of KAGRA due to the molecular adlayer by using the parameters in Table \ref{tab:2}. In this calculation, the formation of a molecular adlayer for the ETM and ITM starts at the same time. Because the original reflectance of the ETM is higher than that of the ITM, the amplitude of reflectance oscillation of the ITM is larger than that of the ETM. The reflectance of ETM decreases over time due to the optical absorption inside the adlayer. On the other hand, the reflectance of ITM does not decrease. This is because the ITM reflectance is lower than the ETM reflectance, and the absorption of the adlayer is not large enough to significantly reduce the ITM reflectance. The reflectance of the arm cavities increases when the reflectance of the ITM decreases because the reflectance of ETM is higher than that of ITM.}\label{TM_ref}
\end{figure}

\begin{figure}[htb]
\centering
\includegraphics[width=8.0cm,clip]{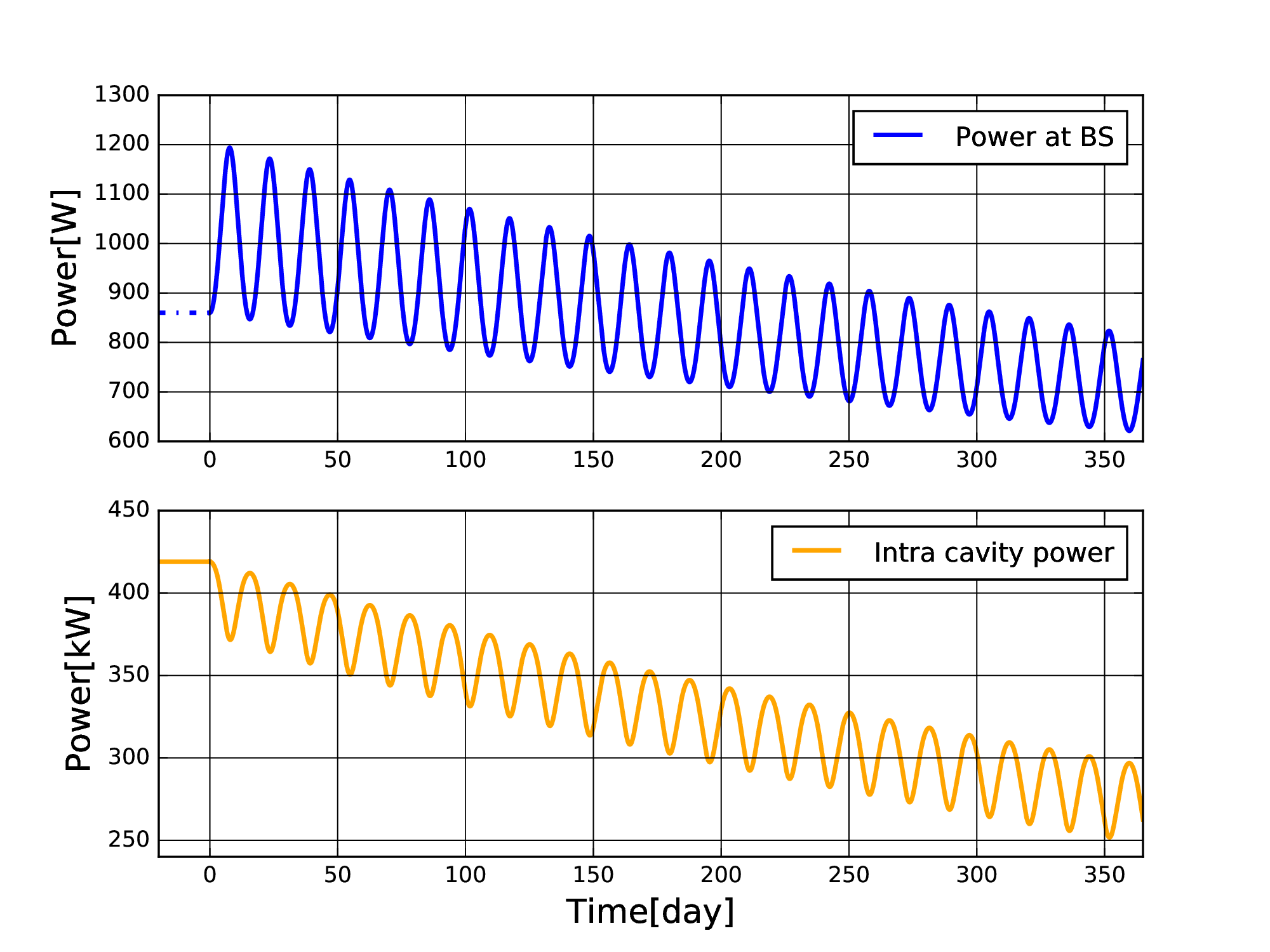}
\caption{Circulating laser power oscillation induced by the adlayer. For this calculation, 80 W of the laser power was assumed as the input power to the power recycling cavity. The reflectance change shown in Fig.\ref{TM_ref} cause the PRG and the finesse oscillation. The changes of these parameters vary the circulating laser power inside the interferometer, especially the laser power at BS and inside the arm cavity. The laser power change at BS directly implies the change of the PRG, and the laser power inside the arm cavity is affected by both PRG and the finesse.}\label{prg}
\end{figure}

\quad It is important to consider how this molecular adlayer affects the sensitivity of KAGRA. We assume that the formation of a molecular adlayer starts at the same time and the accumulation speed is also same. The effect of the molecular desorption is not taken into account in the following estimations.\\
\quad Shot noise and radiation pressure noise, called quantum noise, will limit the sensitivity of KAGRA and next-generation GWDs. A power recycling technique is indispensable to improve the contribution of the shot noise by enhancing the laser power inside the interferometers \cite{Ando1997, Sato2000}. In power recycling techniques, it is important to match the reflectance of a power recycling mirror with that of the arm cavities. A power recycling gain (PRG), which is defined as a ratio between the laser power before and inside a power recycling cavity, is given by 
 \begin{equation}\label{eq:7}
 \mathrm{G}=\left( \frac{ t_{\mathrm{prm}} }{1- \frac{1}{2}r_{\mathrm{prm}} \left(r_{\mathrm{fpx}}+r_{\mathrm{fpy}}   \right) }    \right)^2 ,
 \end{equation}
where $t_{\mathrm{prm}}$ and $r_{\mathrm{prm}}$ are the amplitude transmittance and reflectance of the power recycling mirror, and $r_{\mathrm{fpx}}$ and $r_{\mathrm{fpy}}$ are the amplitude reflectances of each arm cavity. From Eq.(\ref{eq:7}), the power recycling gain is maximized when $r_{\mathrm{prm}}=\left ( r_{\mathrm{fpx}} + r_{\mathrm{fpy}}  \right ) /2 $, and actual GWDs are designed to satisfy this condition. In KAGRA, these parameters are set as $\left | t_{\mathrm{prm}} \right |^2= 0.1 $, $\left | r_{\mathrm{prm}} \right |^2  =  0.9 $, $\left | r_{\mathrm{fpx}} \right | ^2 = \left | r_{\mathrm{fpy}} \right | ^2  = 0.9 $, and $G\sim10$ \cite{Michimura2018}. 
Figure \ref{TM_ref} shows the reduction and oscillation of reflectance of the cryogenic mirrors in KAGRA due to the adlayer growth. In this calculation, the formation of a molecular adlayer for the end test mass (ETM) and the input test mass (ITM) starts at the same time. Because the original reflectance of the ETM is higher than that of the ITM, the amplitude of reflectance oscillation of the ITM is larger than that of the ETM. The reflectance of ETM decreases over time due to the optical absorption inside the adlayer. On the other hand, the reflectance of ITM does not decrease. This is because the ITM reflectance is lower than the ETM reflectance, and the absorption of the adlayer is not large enough to significantly reduce the ITM reflectance. The reflectance of the arm cavities increase when the reflectance of the ITM decreases because the reflectance of ETM is higher than that of ITM.\\
\quad The reflectance oscillation shown in Fig.\ref{TM_ref} cause the PRG and the finesse change. The changes of these parameters vary the circulating laser power inside the interferometer, especially at BS and inside the arm cavity. Fig.\ref{prg} shows the circulating laser power fluctuation at the beam splitter (BS) and inside the arm cavity induced by the PRG and the finesse change. The laser power change at BS directly implies the change of the PRG, and the laser power change inside the arm cavity is induced by both of the PRG and the finesse.
 
% \quad The transferable heat from an object at temperature $T_2$ to an object at temperature $T1$ is given by the Fourier's law :
% \begin{equation}
%Q = \int^{T_2}_{T_1} {\frac{\pi d^2 N}{4 l} \kappa (d,T)dT},
%\end{equation} 
% where $d$,$l$,$N$ and $\kappa$ are diameter, length, number and thermal conductivity of the fiber or the heat links which are attached to the test massed to transfer the injected heat. Then, the temperature of the test masses are calculated using the specific heat of the sapphire $C_\mathrm{sap}$ as $ T_\mathrm{i}= T_0 + (Q_\mathrm{i}-Q)$ 
% Figure.() shows the calculated temperature fluctuation of the test masses. After the 1 year operation, the absorption of adlayer would be about 25 ppm and because of the large circulated power inside the arm avity, the temperature of the test massed go up to about 45 K.

\begin{figure}[htb]
\centering
\includegraphics[width=8.0cm,clip]{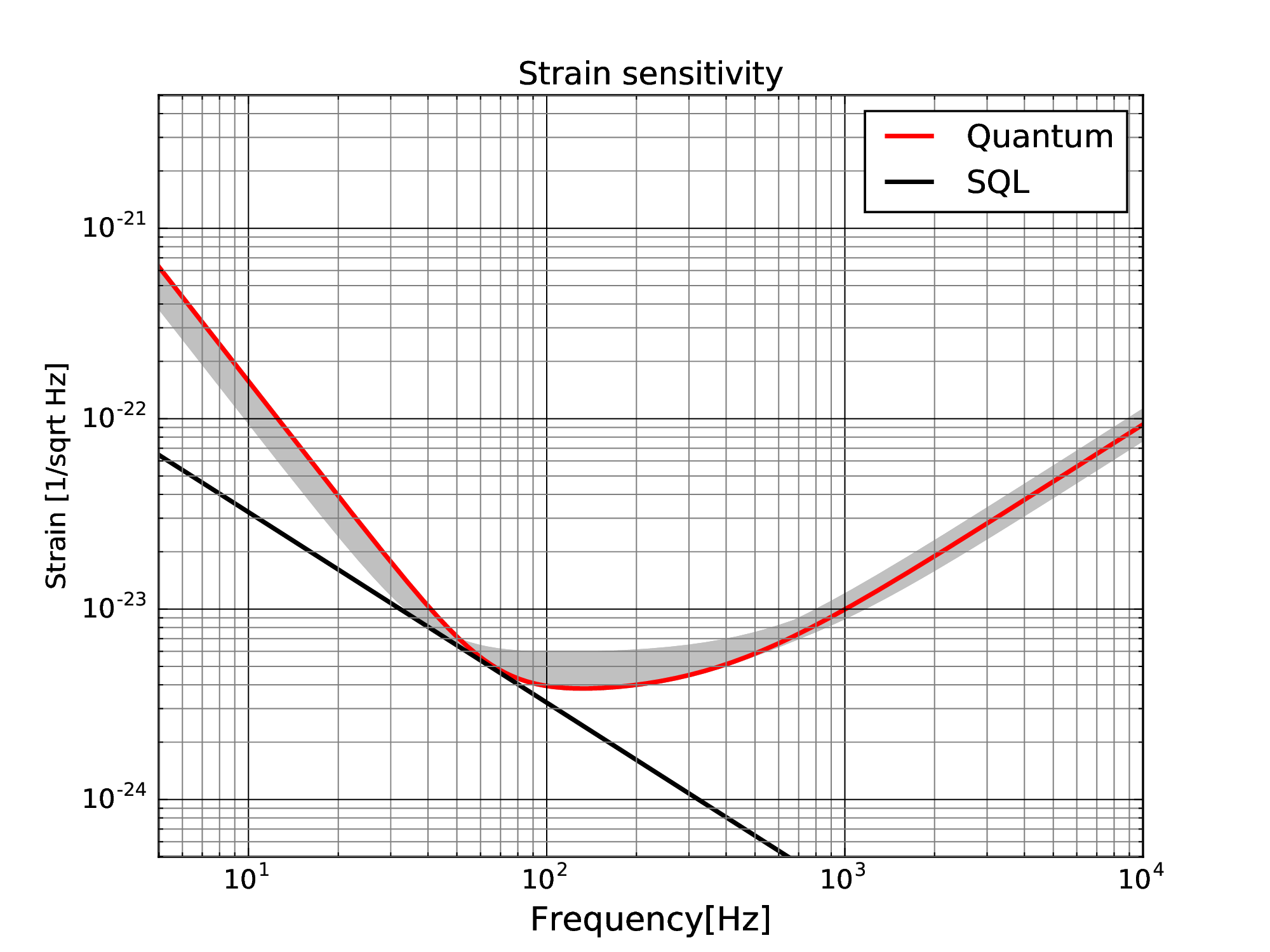}
\caption{The quantum noise in KAGRA. The red solid line shows the designed quantum noise of KAGRA and the gray area shows the quantum noise fluctuation induced by the adlayer on the cryogenic ITM and ETM. The shot noise sometimes improves and at other times gets worse. On the other hand, the radiation pressure noise is improved over time. This is because the shot noise proportional to the laser power at BS, and the radiation pressure noise proportional to the laser power inside the arm cavity.} \label{sensitivity}
\end{figure}

\begin{figure}[htb]
\centering
\includegraphics[width=8.0cm,clip]{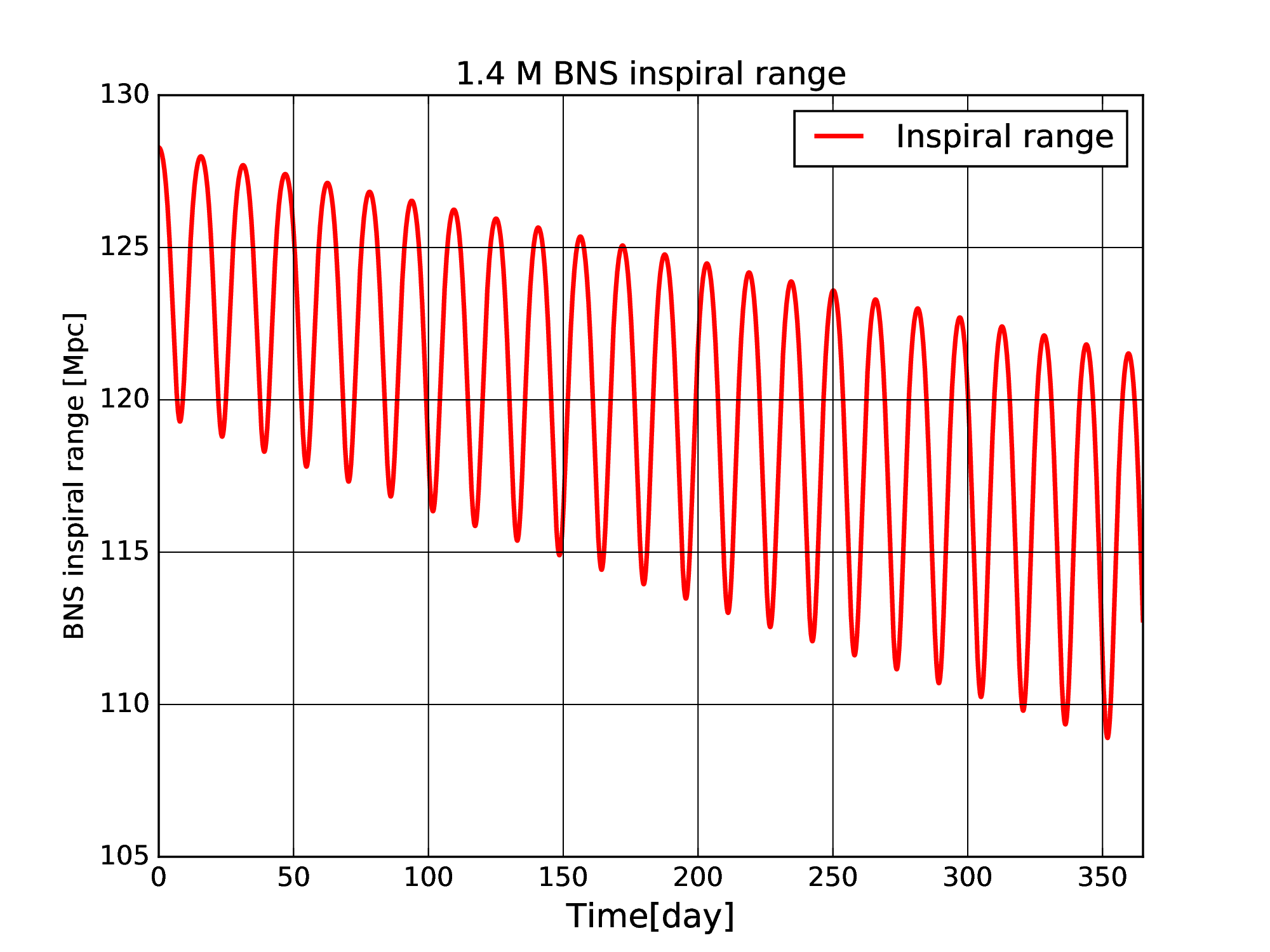}
\caption{The calculated inspiral range of KAGRA. In this calculation, the seismic noise, the mirror and suspension thermal noise and quantum noise are taken into consideration. It is supposed that the fluctuation of the inspiral range is caused only by the quantum noise. Hence, the temperature fluctuation of ITM induced by the circulating laser power is not considered.} \label{range}
\end{figure}

\begin{figure}[htb]
\centering
\includegraphics[width=8.0cm,clip]{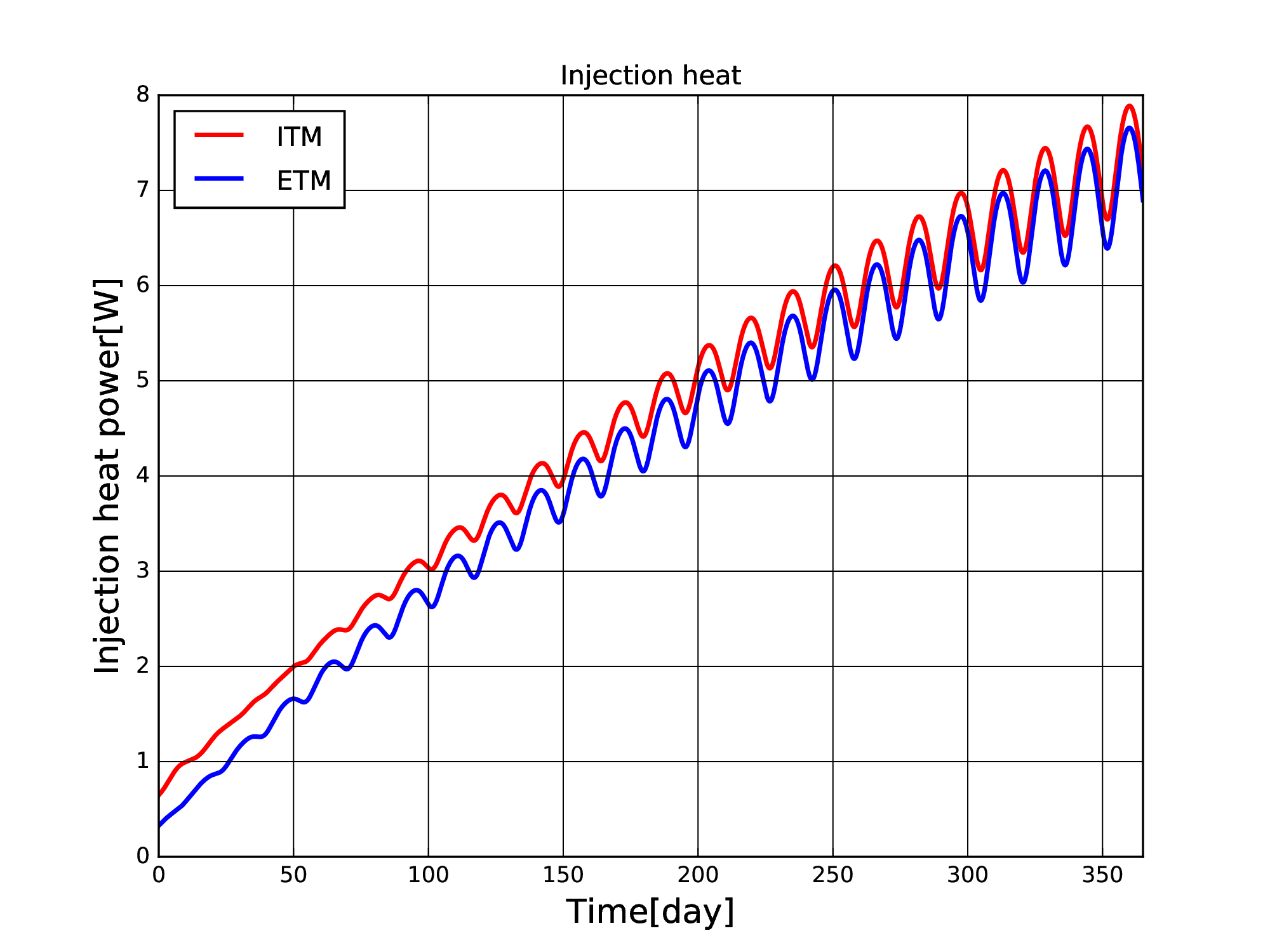}
\caption{The calculated total input heat to the test masses. The radiation from the beam ducts and the absorption of the substrate, the coating, and the adlayer are taken into account. After one year of the operation, the laser power at BS and inside the arm cavity decrease to $P_\mathrm{BS} \sim 700$ W and $P_\mathrm{FP} \sim 275$ kW, respectively. On the other hand, the adlayer grows up to about 10 $\mathrm{\mu m}$ and it would absorb about 25 ppm of the circulating laser power. The input heat to the test masses is considered to be more than $Q_{\mathrm{ITM}} \sim Q_{\mathrm{ETM}}\sim 7 $ W.} \label{input_heat}
\end{figure}

  The quantum noise of a GWD without a signal recycling cavity is given by 
\begin{equation}\label{eq:shot}
S_{\mathrm{quantum}} \left ( \omega , t  \right ) =  \frac{ 4 \hbar  }{ m \omega ^2 L ^2 } \left ( \frac{1}{\mathcal{K}(t)} + \mathcal{K}(t) \right ) ,
\end{equation}
 where
\begin{equation}
\mathcal{K} (t) = \frac{16\pi c I_0 (t) }{m \lambda L^2 \omega^2 (\gamma^2(t)+\omega^2)}.
\end{equation} 
Here, $\hbar$, $I_0$ and $L$ are Dirac's constant, the input power to a BS and the length of the arm cavity, respectively. $\gamma$ is an arm cavity $\frac{1}{2}$-bandwidth, which is given by \cite{Kimble2001}
\begin{equation}
\gamma (t) = \frac{c T_{\mathrm{ITM}} (t) }{4 L},
\end{equation} 
 where $T_{\mathrm{ITM}}$ is the power transmittance of the ITM. For the actual quantum noise calculation with a signal recycling cavity, Eq.(5.13) in Ref.\cite{Buonanno2001} was applied. Figure \ref{sensitivity} shows the quantum noise calculated using the parameters in Ref.\cite{Michimura2018}. The red solid line shows the designed quantum noise of KAGRA and the gray area shows the quantum noise fluctuation induced by the adlayer on the cryogenic ITM and ETM. The shot noise sometimes improves and at other times gets worse. On the other hand, the radiation pressure noise is improved over time. This is because the shot noise is determined by the laser power at BS, but the radiation pressure noise is determined by the laser power inside the arm cavity.\\
\quad As a result of the quantum noise fluctuation, the inspiral range is also changed. The inspiral range of KAGRA for $1.4M_\odot$ binary neutron star system is calculated using Eq.(19) in Ref.\cite{Michimura2018}. Figure \ref{range} shows the calculated inspiral range of KAGRA. In this calculation, the seismic noise, the mirror and suspension thermal noise and quantum noise are taken into consideration. Here we assume that the molecular adlayer affects only the quantum noise, namely the fluctuation of the inspiral range is caused only by the quantum noise. Before the adlayer formation, the inspiral range is about 127 Mpc with signal recycling. However, during the operation, inspiral range worsens gradually and decreases to about 115 Mpc on average. In other words, the detection rate of gravitational waves possibly decreases about 30\% on average after 1 year operation.
\\
\quad At last, it is meaningful to mention the impact of the adlayer on the optical absorption. The changes of the several parameters could change the temperature of ITMs and ETMs because of the absorption of the substrate, the coating, and the adlayer. The absorption of the substrate $\sigma_{\mathrm{subs}}$ and the coating $\sigma_{\mathrm{coat}}$ is assumed about $\sigma_{\mathrm{subs}} = 750 \mathrm{ppm}$ and $ \sigma_{\mathrm{coat}} = 0.3 \mathrm{ppm}$ at the 1064 nm of the wavelength \cite{Somiya2012}, respectively. The absorption coefficient of the adlayer $\alpha$ is defined as:
 \begin{equation}
\alpha = 4\pi \frac{\mathrm{Im}(N)} {\lambda} ,
\end{equation} 
 where N is the complex refractive index of the adlayer. It is calculated using the measured imaginary part of refractive index as $\alpha =2.60    \ \mathrm{1/m}$. The laser power $I$ inside the adlayer at distance z from the surface is described as $I(z)=I_0 \exp{\left ( -\alpha z \right )}$, and the absorption of the adlayer $\sigma_\mathrm{ad}$ is defined as : 
\begin{equation}
\sigma_{\mathrm{ad}} = (1-\exp{\left ( -\alpha z \right ))}.
\end{equation}   
    The input heat to the ITM $Q_{\mathrm{ITM}}$ and the ETM $Q_\mathrm{ETM}$ due to the light absorption are approximated as :
\begin{equation}\label{eq:input_heat_1}
Q_\mathrm{ITM}    \sim  \frac{P_\mathrm{BS}}{2}  \left ( \sigma_{\mathrm{subs}}+\sigma_\mathrm{ad} \right ) + P_{\mathrm{FP}} \left ( \sigma_{\mathrm{coat}}+\sigma_\mathrm{ad}  \right ),
\end{equation}
\begin{equation}\label{eq:input_heat_2}
Q_\mathrm{ETM}    \sim P_{\mathrm{FP}} \left ( \sigma_{\mathrm{coat}}+\sigma_\mathrm{ad}  \right ),
\end{equation}  
where $P_\mathrm{BS}$ and $P_\mathrm{FP}$ are the circulating laser power at the BS and inside the arm cavity. In the Eq.(\ref{eq:input_heat_1}) and (\ref{eq:input_heat_2}), we ignored the absorption of  the substrate for the ETM and the absorption of the antireflection coating for both of test masses. Figure \ref{input_heat} shows the calculated total input heat to the test masses. The radiation from the beam ducts and the absorption of the substrate, the coating, and the adlayer are taken into account. After one year of the operation, the laser power at BS and inside the arm cavity decrease to $P_\mathrm{BS} \sim 700$ W and $P_\mathrm{FP} \sim 275$ kW, respectively. On the other hand, the adlayer grows up to about 10 $\mathrm{\mu m}$ and it would absorb about 25 ppm of the circulating laser power. The input heat to the test masses is considered to be more than $Q_{\mathrm{ITM}} \sim Q_{\mathrm{ETM}}\sim 7 $ W. These large heat injection warm the test masses up. When the temperature of the test mass increase, the sticking probability of molecules to the test mass decrease and the number of molecules which get out from the adlayer increase. Finally, it would be the adsorption-desorption equilibrium state. Then, the adlayer growing will stop and the input heat to the test mass will not increase any more. In order to determine the desorption rate and the thickness of the adlayer at the equilibrium state, the measurement of the surface potential depth is required.\\
\quad The next generation GWDs are focusing on the longer wavelength of the laser, such as 1550 nm and 2000 nm. The absorption of the LDA state of water molecules at these wavelengths is considered to be much higher than 1064 nm of the laser \cite{Warren2008}. The absorption of the adlayer can be a big problem. On the other hand, the higher temperature of the test mass, such as 120K, can decrease the thickness of the adlayer at the adsorption-desorption equilibrium state.\\
\quad From the point of the view of the thermal noise, the calculation of the thermal response of the suspension system to the heat injection for the test mass have to be studied. Furthermore, the adlayer itself can be a source of the thermo-optic noise \cite{Evans2008}. The multilateral study is required for the understanding of the impacts of the adlayer on the cryogenic GWDs.\\

% Figure \ref{shot_noise} shows the relative shot noise $( \sqrt{ S_{\mathrm{shot}} \left ( \omega , t  \right )/ S_{\mathrm{shot}} \left ( \omega , 0  \right ) }) $ of KAGRA at 100 Hz. The shot noise changes in phase with the reflectance of the arm cavities and PRG. As a result of the molecular adlayer formation, the shot noise is improved by approximately 7$\%$ on average at first. But during the operation, the shot noise changes gradually, and it is expected that the shot noise worsens by more than 10$\%$ on average after one year of operation, owing to the optical loss inside. 
 
% \begin{figure}[htb]
%\centering
%\includegraphics[width=8.0cm,clip]{shot_noise_ratio.eps}
%\caption{Result of calculation of relative shot noise in KAGRA at 100 Hz. Using Eq.(\ref{eq:shot}), the ratio of the shot noise $( \sqrt{ S_{\mathrm{shot}} \left ( \omega , t  \right )/ S_{\mathrm{shot}} \left ( \omega , 0  \right ) }) $ was calculated at 100 Hz. At first, the shot noise will decrease by more than 7$\%$ on average because the circulating power inside the interferometer increases owing to the PRG enhancement. But after one year of operation, it gets worse by more than 10$\%$ on average, owing to the optical loss inside the molecular adlayer.}\label{shot_noise}
%\end{figure}
% Furthermore, this molecular adlayer possibly causes many problems in cryogenic GWDs. For example, the thermo-optic noise that is the noise induced by the temperature dependence of the refractive inde  such as the thermal noise and the difficulty of controls. 

%%%%%%%%%%%%%%%%%%%%%%%%%%%%%%%%%%%%%%%%%%%%%%%%%%%%%%%%%%%
\section{Conclusion}
We measured the molecular adlayer formation on a cold mirror using a high-finesse optical cavity in a KAGRA cryostat. The adlayer formation speed was evaluated by simulation and an experiment. Although there is a difference between the expectation from the simulation and the experimental results, the difference could arise from the uncertainty of the vacuum level setting or inhomogeneous outgassing. The effect of a molecular adlayer on the optical coating was discussed based on the thin film coating theory and the experimental results. A molecular adlayer will change the mirror reflectance and it causes the circulating laser power fluctuation inside the interferometer. The fluctuation of these parameters cause the quantum noise fluctuation, and finally, the inspiral range for the 1.4$M_\odot$ binary neutron star system worsens about 30\% on average after one year of operation. Furthermore, this adlayer possibly causes other significant problems, such as worse thermal noise and difficulty of control of interferometer. The characterization of the molecular desorption is much important to estimate the impacts of the adlayer on the cryogenic GWDs.\\
\quad To compensate for the molecular accumulation and to remove the accumulated molecules on a cold mirror, some methods were discussed. These methods need further measurements and study to confirm their efficiency and stability. In order to archive high sensitivity and stable operation with cryogenic GWDs, these types of problems have to be solved.

 %%%%%%%%%%%%%%%%%%%%%%%%%%%%%%%%%%%%%%%%%%%%%%%%%%%%%%%%%%%
\begin{acknowledgments}
We thank Dr. Y. Inoue of Academia Sinica in Taiwan for his kind help. We acknowledge financial support for K.H. from the Advanced Leading Graduate Course for Photon Science (ALPS) program at the University of Tokyo. This work was supported by the Center for the Promotion of Integrated Sciences (CPIS), SOKENDAI (The Graduate University for Advanced Studies), MEXT, JSPS Leading-edge Research Infrastructure Program, JSPS Grant-in-Aid for Specially Promoted Research 26000005, JSPS Grant-in-Aid for Scientific Research on Innovative Areas 2905: JP17H06358, JP17H06361 and JP17H06364, JSPS Core-to-Core Program A. Advanced Research Networks, JSPS Grant-in-Aid for Scientific Research (S) 17H06133, the joint research program of the Institute for Cosmic Ray Research, University of Tokyo, National Research Foundation (NRF) and Computing Infrastructure Project of KISTI-GSDC in Korea, the LIGO project, and the Virgo project.
\end{acknowledgments}

%\nocite{*}
%\bibliography{PRD_Reference.bib}% Produces the bibliography via BibTeX.




%\bibliography{PRD_Reference.bib}

\begin{thebibliography}{28}%
\makeatletter
\providecommand \@ifxundefined [1]{%
 \@ifx{#1\undefined}
}%
\providecommand \@ifnum [1]{%
 \ifnum #1\expandafter \@firstoftwo
 \else \expandafter \@secondoftwo
 \fi
}%
\providecommand \@ifx [1]{%
 \ifx #1\expandafter \@firstoftwo
 \else \expandafter \@secondoftwo
 \fi
}%
\providecommand \natexlab [1]{#1}%
\providecommand \enquote  [1]{``#1''}%
\providecommand \bibnamefont  [1]{#1}%
\providecommand \bibfnamefont [1]{#1}%
\providecommand \citenamefont [1]{#1}%
\providecommand \href@noop [0]{\@secondoftwo}%
\providecommand \href [0]{\begingroup \@sanitize@url \@href}%
\providecommand \@href[1]{\@@startlink{#1}\@@href}%
\providecommand \@@href[1]{\endgroup#1\@@endlink}%
\providecommand \@sanitize@url [0]{\catcode `\\12\catcode `\$12\catcode
  `\&12\catcode `\#12\catcode `\^12\catcode `\_12\catcode `\%12\relax}%
\providecommand \@@startlink[1]{}%
\providecommand \@@endlink[0]{}%
\providecommand \url  [0]{\begingroup\@sanitize@url \@url }%
\providecommand \@url [1]{\endgroup\@href {#1}{\urlprefix }}%
\providecommand \urlprefix  [0]{URL }%
\providecommand \Eprint [0]{\href }%
\providecommand \doibase [0]{http://dx.doi.org/}%
\providecommand \selectlanguage [0]{\@gobble}%
\providecommand \bibinfo  [0]{\@secondoftwo}%
\providecommand \bibfield  [0]{\@secondoftwo}%
\providecommand \translation [1]{[#1]}%
\providecommand \BibitemOpen [0]{}%
\providecommand \bibitemStop [0]{}%
\providecommand \bibitemNoStop [0]{.\EOS\space}%
\providecommand \EOS [0]{\spacefactor3000\relax}%
\providecommand \BibitemShut  [1]{\csname bibitem#1\endcsname}%
\let\auto@bib@innerbib\@empty
%</preamble>
\bibitem [{\citenamefont {Abbott}(2016)}]{Abbott2016}%
  \BibitemOpen
  \bibfield  {author} {\bibinfo {author} {\bibfnamefont {B. P. Abbott {\it et al.}}},\ }
  \href {\doibase 10.1103/PhysRevLett.116.061102} {\bibfield
  {journal} {\bibinfo  {journal} {Physical Review Letters}\ }\textbf {\bibinfo
  {volume} {061102}},\ \bibinfo {pages} {1} (\bibinfo {year}
  {2016})}\BibitemShut {NoStop}%
  \bibitem [{\citenamefont {Abbott}(2017)}]{Abbott2017a}%
  \BibitemOpen
  \bibfield  {author} {\bibinfo {author} {\bibfnamefont {B. P. Abbott {\it et al.}}},\ }\href {\doibase
  10.1103/PhysRevLett.119.141101} {\bibfield  {journal} {\bibinfo  {journal}
  {Physical Review Letters}\ }\textbf {\bibinfo {volume} {119}},\ \bibinfo
  {pages} {1} (\bibinfo {year} {2017})}\BibitemShut {NoStop}%  
  \bibitem [{\citenamefont {Kinugawa}(2014)}]{Kinugawa2014}%
  \BibitemOpen
  \bibfield  {author} {\bibinfo {author} {\bibfnamefont {T. Kinugawa {\it et al.}}},\ }\href {\doibase
  10.1093/mnras/stu1022} {\bibfield  {journal} {\bibinfo  {journal} {MNRAS}\
  }\textbf {\bibinfo {volume} {442}},\ \bibinfo {pages} {2963} (\bibinfo {year}
  {2014})}\BibitemShut {NoStop}%
  \bibitem [{\citenamefont {Harry}(2007)}]{Harry2007}%
  \BibitemOpen
  \bibfield  {author} {\bibinfo {author} {\bibfnamefont {G. M. Harry {\it et al.}}},\ }\href {\doibase 
 10.1088/0264-9381/24/2/008} {\bibfield
  {journal} {\bibinfo  {journal} {Classical and Quantum Gravity}\ }\textbf
  {\bibinfo {volume} {24}},\ \bibinfo {pages} {405} (\bibinfo {year}
  {2007})},\ \Eprint {https://arxiv.org/abs/gr-qc/0610004}{arxiv:gr-qc/0610004v1} \BibitemShut {NoStop}%
  \bibitem [{\citenamefont {Cole}(2013)}]{Cole2013}%
  \BibitemOpen
  \bibfield  {author} {\bibinfo {author} {\bibfnamefont {G. D. Cole {\it et al.}}},\ }\href {\doibase
  10.1038/nphoton.2013.174} {\bibfield  {journal} {\bibinfo  {journal} {Nature
  Photonics}\ }\textbf {\bibinfo {volume} {7}},\ \bibinfo {pages} {644}
  (\bibinfo {year} {2013})}\BibitemShut {NoStop}%
  \bibitem [{\citenamefont {Mours}(2006)}]{Mours2006}%
  \BibitemOpen
  \bibfield  {author} {\bibinfo {author} {\bibfnamefont {B. Mours {\it et al.}}},\ }\href {\doibase 
  10.1088/0264-9381/23/20/001} {\bibfield  {journal} {\bibinfo
  {journal} {Classical and Quantum Gravity}\ }\textbf {\bibinfo {volume}
  {23}},\ \bibinfo {pages} {5777} (\bibinfo {year} {2006})}\BibitemShut{NoStop}%
\bibitem [{\citenamefont {Somiya}(2012)}]{Somiya2012}%
  \BibitemOpen
  \bibfield  {author} {\bibinfo {author} {\bibfnamefont {K. Somiya {\it et al.}}},\ }\href {\doibase 
  10.1088/0264-9381/29/12/124007} {\bibfield
  {journal} {\bibinfo  {journal} {Classical and Quantum Gravity}\ }\textbf
  {\bibinfo {volume} {29}},\ \bibinfo {pages} {124007} (\bibinfo {year}
  {2012})},\ \Eprint {http://arxiv.org/abs/1111.7185} {arXiv:1111.7185}\BibitemShut {NoStop}%
\bibitem [{\citenamefont {Aso}(2013)}]{Aso2013}%
  \BibitemOpen
  \bibfield  {author} {\bibinfo {author} {\bibfnamefont {Y. Aso {\it et al.}}},\ }\href {\doibase
  10.1103/PhysRevD.88.043007} {\bibfield  {journal} {\bibinfo  {journal}
  {Physical Review D}\ }\textbf {\bibinfo {volume} {88}},\ \bibinfo {pages} {1}
  (\bibinfo {year} {2013})},\ \Eprint {http://arxiv.org/abs/1306.6747}
  {arXiv:1306.6747} \BibitemShut {NoStop}%
  \bibitem [{\citenamefont {Khalaidovski}(2014)}]{Khalaidovski2014}%
  \BibitemOpen
  \bibfield  {author} {\bibinfo {author} {\bibfnamefont {A.Khalaidovski {\it et al.}}},\ }\href {\doibase doi:10.1088/0264-9381/31/10/105004}{\bibfield  {journal} {\bibinfo  {journal}  {Class. Quantum Grav.}\ }\textbf {\bibinfo {volume} {31}},\ \bibinfo {pages} {105004}
  (\bibinfo {year} {1999})}\BibitemShut {NoStop}%  
  \bibitem [{\citenamefont {Uchiyama}(1999)}]{Uchiyama1999}%
  \BibitemOpen
  \bibfield  {author} {\bibinfo {author} {\bibfnamefont {T.Uchiyama {\it et al.}}},\ }\href {\doibase 10.1016/S0375-9601(99)00563-0}{\bibfield  {journal} {\bibinfo  {journal}  {Physics Letters A}\ }\textbf {\bibinfo {volume} {261}},\ \bibinfo {pages} {5-11}
  (\bibinfo {year} {1999})}\BibitemShut {NoStop}%  
    \bibitem [{\citenamefont {P.Amico}(2004)}]{Amico2004}%
  \BibitemOpen
  \bibfield  {author} {\bibinfo {author} {\bibfnamefont {P.Amico {\it et al.}}},\ }\href {\doibase 10.1016/j.nima.2003.10.071}{\bibfield  {journal} {\bibinfo  {journal}  {Nucl. Instrum. Methods Phys. Res. A}\ }\textbf {\bibinfo {volume} {518}},\ \bibinfo {pages} {240-243}
  (\bibinfo {year} {2004})}\BibitemShut {NoStop}%  
    \bibitem [{\citenamefont {R.Nawrodt}(2008)}]{Nawrodt2008}%
  \BibitemOpen
  \bibfield  {author} {\bibinfo {author} {\bibfnamefont {R.Nawrodt {\it et al.}}},\ }\href {\doibase 10.1088/1742-6596/122/1/012008}{\bibfield  {journal} {\bibinfo  {journal}  {J.Phys.:Conf. Ser.}\ }\textbf {\bibinfo {volume} {122}},\ \bibinfo {pages} {012008}
  (\bibinfo {year} {1999})}\BibitemShut {NoStop}%   
\bibitem [{\citenamefont {Punturo}(2010)}]{Punturo2010}%
  \BibitemOpen
  \bibfield  {author} {\bibinfo {author} {\bibfnamefont {M. Punturo {\it et al.}}},\ }\href {\doibase
   10.1088/0264-9381/27/19/194002}
  {\bibfield  {journal} {\bibinfo  {journal} {Classical and Quantum Gravity}\
  }\textbf {\bibinfo {volume} {27}} (\bibinfo {year} {2010})}\BibitemShut {NoStop}%
\bibitem [{\citenamefont {Hild}(2012)}]{Hild2012}%
  \BibitemOpen
  \bibfield  {author} {\bibinfo {author} {\bibfnamefont {S. Hild {\it et al.}}},\ }\href {\doibase
  10.1088/0264-9381/29/12/124006} {\bibfield
  {journal} {\bibinfo  {journal} {Classical and Quantum Gravity}\ }\textbf
  {\bibinfo {volume} {29}} (\bibinfo {year} {2012})},\ \Eprint {http://arxiv.org/abs/1111.6277}
  {arXiv:1111.6277} \BibitemShut {NoStop}%
\bibitem [{\citenamefont {Tomaru}(2001)}]{Tomaru2001}%
  \BibitemOpen
  \bibfield  {author} {\bibinfo {author} {\bibfnamefont {T. Tomaru {\it et al.}}},\ }\href {\doibase
  10.1016/S0375-9601(01)00191-8} {\bibfield  {journal} {\bibinfo  {journal}
  {Physics Letters, Section A: General, Atomic and Solid State Physics}\
  }\textbf {\bibinfo {volume} {283}},\ \bibinfo {pages} {80} (\bibinfo {year}
  {2001})}\BibitemShut {NoStop}%
\bibitem [{\citenamefont {Uchiyamaa}(1998)}]{Uchiyama1998}%
  \BibitemOpen
  \bibfield  {author} {\bibinfo {author} {\bibfnamefont {T. Uchiyamaa {\it et al.}}},\
  }\href@noop {} {\bibfield  {journal} {\bibinfo  {journal} {Physics Letters
  A}\ }\textbf {\bibinfo {volume} {242}},\ \bibinfo {pages} {211} (\bibinfo
  {year} {1998})}\BibitemShut {NoStop}%
\bibitem [{\citenamefont {Tomaru}(2004)}]{Tomaru2004}%
  \BibitemOpen
  \bibfield  {author} {\bibinfo {author} {\bibfnamefont {T. Tomaru {\it et al.}}},\
  }\href@noop {} {\bibfield  {journal} {\bibinfo  {journal} {Classical and Quantum Gravity}\ }\textbf {\bibinfo {volume} {21}},\ \bibinfo {pages} {1005} (\bibinfo
  {year} {2004})}\BibitemShut {NoStop}%
    \bibitem [{\citenamefont {Tokoku}(2014)}]{Tokoku2014}%
  \BibitemOpen
  \bibfield  {author} {\bibinfo {author} {\bibfnamefont {C. Tokoku {\it et al.}}},\ }\href {\doibase
  10.1063/1.4860850} {\bibfield
  {journal} {\bibinfo  {journal} {AIP Conference Proceedings}\ }\textbf
  {\bibinfo {volume} {1573}},\ \bibinfo {pages} {1254} (\bibinfo {year}
  {2014})}\BibitemShut {NoStop}%  
    \bibitem [{\citenamefont {Warren}(2008)}]{Warren2008}%
  \BibitemOpen
  \bibfield  {author} {\bibinfo {author} {\bibfnamefont {S. G. Warren and R. E. Brandt}},\ }\href {\doibase doi:1029/2007JD009744}{\bibfield  {journal} {\bibinfo  {journal}  {Jour. Geophys. Res.}\ }\textbf {\bibinfo {volume} {113}},\ \bibinfo {pages} {458}
  (\bibinfo {year} {2008})}\BibitemShut {NoStop}%  
  \bibitem [{\citenamefont {Sakakibara}(2015)}]{Sakakibara2015}%
  \BibitemOpen
  \bibfield  {author} {\bibinfo {author} {\bibfnamefont {Y. Sakakibara {\it et al.}}},\
  }\href@noop {} {\bibfield  {journal} {\bibinfo  {journal} {Classical and Quantum Gravity}\ }\textbf {\bibinfo {volume} {32}},\ \bibinfo {pages} {155011} (\bibinfo
  {year} {2015})}\BibitemShut {NoStop}%
    \bibitem [{\citenamefont {Miyoki}(2001)}]{Miyoki2001}%
  \BibitemOpen
  \bibfield  {author} {\bibinfo {author} {\bibfnamefont {S. Miyoki {\it et al.}}},\
  }\href@noop {} {\bibfield  {journal} {\bibinfo  {journal} {Cryogenics}\ }\textbf {\bibinfo {volume} {41}},\ \bibinfo {pages} {415-420} (\bibinfo
  {year} {2001})}\BibitemShut {NoStop}%
  \bibitem [{\citenamefont {Takahashi}(2009)}]{Takahashi2009}%
  \BibitemOpen
  \bibfield  {author} {\bibinfo {author} {\bibfnamefont {R. Takahashi and Y. Saito}},\ }{\bibfield
  {journal} {\bibinfo  {journal} {Vacuum}\ }\textbf
  {\bibinfo {volume} {84}},\ \bibinfo {pages} {709-912} (\bibinfo {year}
  {2009})}  \BibitemShut {NoStop}%
  \bibitem [{\citenamefont {Marquardt}(1999)}]{Marquardt1999}%
  \BibitemOpen
  \bibfield  {author} {\bibinfo {author} {\bibfnamefont {N. Marquardt}},\ }\href {\doibase
   10.5170/CERN-1999-005.1} {\bibfield
  {journal} {\bibinfo  {journal} {Vacuum}\ \bibinfo {pages} {1}} (\bibinfo
  {year} {1999})}\BibitemShut {NoStop}%
\bibitem [{\citenamefont {Lobo}(2004)}]{Lobo2004}%
  \BibitemOpen
  \bibfield  {author} {\bibinfo {author} {\bibfnamefont {P. J. Lobo, F. Becheri and J. {G{\'{o}}mez-Go{\~{n}}i}}},\ }\href {\doibase 
  10.1016/j.vacuum.2004.05.013}
  {\bibfield  {journal} {\bibinfo  {journal} {Vacuum}\ }\textbf {\bibinfo
  {volume} {76}},\ \bibinfo {pages} {83} (\bibinfo {year} {2004})}\BibitemShut
  {NoStop}%
  \bibitem [{\citenamefont {Corruccini}(1961)}]{Corruccini1961}%
  \BibitemOpen
  \bibfield  {author} {\bibinfo {author} {\bibfnamefont { R. J. Corruccini and J. J. Gniewek}},\ }\href@noop {} {\bibfield  {journal} {\bibinfo  {journal} { NBS Monograph 29, National Bureau of Standards, Boulder, CO}\ } (\bibinfo {year} {1961})}\BibitemShut
  {NoStop}%
  \bibitem [{\citenamefont {Labello}(2011)}]{Labello2011}%
  \BibitemOpen
  \bibfield  {author} {\bibinfo {author} {\bibfnamefont {J. M. Labello}},\ }\href@noop {} {\bibfield  {journal} {\bibinfo  {journal} {PhD
  Thesis, University of Tennessee}\ } (\bibinfo {year} {2011})}\BibitemShut
  {NoStop}%
\bibitem [{\citenamefont {Westley}(1998)}]{Westleya1998}%
  \BibitemOpen
  \bibfield  {author} {\bibinfo {author} {\bibfnamefont {M. S. Westley, G. A. Baratta and R. A. Baragiola}},\ }\href@noop {} {\bibfield  {journal} {\bibinfo  {journal}
  {Journal of Chemical Physics}\ }\textbf {\bibinfo {volume} {108}} (\bibinfo
  {year} {1998})}\BibitemShut {NoStop}%
  \bibitem{Molflow}Molflow+ : \url{https://molflow.web.cern.ch/}
  \bibitem [{\citenamefont {Focarino}(1990)}]{John1990}%
  \BibitemOpen
  \bibfield  {author} {\bibinfo {author} {\bibfnamefont {John L. Emmett, Pleasanton and Calif}},\ }\href@noop {} {\bibfield  {journal} {\bibinfo
  {journal} {United States Patent}\ {4,925,259} } (\bibinfo {year} {1990})}\BibitemShut
  {NoStop}%
  \bibitem [{\citenamefont {Paul}(2001)}]{Paul2001}%
  \BibitemOpen
  \bibfield  {author} {\bibinfo {author} {\bibfnamefont {J. B. Paul, L. Lapson and J. G. Anderson}},\ }\href
  {\doibase 10.1364/AO.40.004904} {\bibfield  {journal} {\bibinfo  {journal}
  {Applied optics}\ }\textbf {\bibinfo {volume} {40}},\ \bibinfo {pages} {4904}
  (\bibinfo {year} {2001})}\BibitemShut {NoStop}%
\bibitem [{\citenamefont {Uehara}(1995)}]{Uehara1995a}%
  \BibitemOpen
  \bibfield  {author} {\bibinfo {author} {\bibfnamefont {N. Uehara {\it et al.}}},\ }\href {\doibase
   10.1364/OL.20.000530} {\bibfield  {journal}
  {\bibinfo  {journal} {Optics Letters}\ }\textbf {\bibinfo {volume} {20}},\
  \bibinfo {pages} {530} (\bibinfo {year} {1995})}\BibitemShut {NoStop}%
\bibitem [{\citenamefont {Uehara}(1995)}]{Uehara1995}%
  \BibitemOpen
  \bibfield  {author} {\bibinfo {author} {\bibfnamefont {N. Uehara and K. Ueda}},\ }\href {\doibase 
  10.1007/BF01090966} {\bibfield  {journal} {\bibinfo
  {journal} {Applied optics B}\ }\textbf {\bibinfo {volume} {15}},\ \bibinfo
  {pages} {9} (\bibinfo {year} {1995})}\BibitemShut {NoStop}%
\bibitem [{\citenamefont {Isogai}(2013)}]{Isogai2013a}%
  \BibitemOpen
  \bibfield  {author} {\bibinfo {author} {\bibfnamefont {T. Isogai {\it et al.}}},\ }\href {\doibase
  10.1364/OE.21.030114} {\bibfield  {journal} {\bibinfo  {journal} {Optics
  express}\ }\textbf {\bibinfo {volume} {21}},\ \bibinfo {pages} {30114}
  (\bibinfo {year} {2013})},\ \Eprint {http://arxiv.org/abs/arXiv:1310.1820v2}
  {arXiv:arXiv:1310.1820v2} \BibitemShut {NoStop}%
  \bibitem [{\citenamefont {Hendricks}(2015)}]{Hendricks2015}%
  \BibitemOpen
  \bibfield  {author} {\bibinfo {author} {\bibfnamefont {Jay H. Hendricks {\it et al.}}},\ } {\bibfield
  {journal} {\bibinfo  {journal} {IMEKO XXI World Congress ``Measurement in Research and Industry"}\ }\textbf
  {\bibinfo {volume} {1}},\ \bibinfo {pages} {1574} (\bibinfo {year}
  {2015})}\BibitemShut {NoStop}%
  \bibitem [{\citenamefont {Vinet}(1996)}]{Vinet1996}%
  \BibitemOpen
  \bibfield  {author} {\bibinfo {author} {\bibfnamefont {J.-Y. Vinet, V. Brisson and S. Braccini}},\ } {\bibfield
  {journal} {\bibinfo  {journal} {Physical Review D}\ }
  \textbf{\bibinfo {volume} {54}},\ \bibinfo {pages} {1276} 
  (\bibinfo {year}
  {1996})}\BibitemShut {NoStop}%
  \bibitem [{\citenamefont {Zeidler}(2017)}]{Zeidler2017}%
  \BibitemOpen
  \bibfield  {author} {\bibinfo {author} {\bibfnamefont {S. Zeidler {\it et al.}}},\ } {\bibfield
  {journal} {\bibinfo  {journal} {Optics Express}\ }\textbf
  {\bibinfo {volume} {25}},\ \bibinfo {pages} {4741} (\bibinfo {year}
  {2017})}\BibitemShut {NoStop}%
  \bibitem [{\citenamefont {Focsa}(2006)}]{Focsa2006}%
  \BibitemOpen
  \bibfield  {author} {\bibinfo {author} {\bibfnamefont {C. Focsa {\it et al.}}},\ }\href {\doibase
  10.1088/0953-8984/18/30/S02} {\bibfield  {journal} {\bibinfo  {journal}
  {Journal of Physics Condensed Matter}\ }\textbf {\bibinfo {volume} {18}}
  (\bibinfo {year} {2006}),\ 10.1088/0953-8984/18/30/S02}\BibitemShut {NoStop}%
\bibitem [{\citenamefont {Fraser}(2001)}]{Fraser2001}%
  \BibitemOpen
  \bibfield  {author} {\bibinfo {author} {\bibfnamefont {H. J. Fraser {\it et al.}}},\ }\href {\doibase 
  10.1046/j.1365-8711.2001.04835.x} {\bibfield
  {journal} {\bibinfo  {journal} {Monthly Notices of the Royal Astronomical
  Society}\ }\textbf {\bibinfo {volume} {327}},\ \bibinfo {pages} {9} (\bibinfo
  {year} {2001})},\ \Eprint {http://arxiv.org/abs/0107487} {arXiv:0107487
  [astro-ph]} \BibitemShut {NoStop}%
\bibitem [{\citenamefont {Ando}(1997)}]{Ando1997}%
  \BibitemOpen
  \bibfield  {author} {\bibinfo {author} {\bibfnamefont {M. Ando, K. Kawabe and K. Tsubono}},\ }\href
  {\doibase 10.1016/S0375-9601(97)00745-7} {\bibfield  {journal} {\bibinfo
  {journal} {Physics Letters A}\ }\textbf {\bibinfo {volume} {237}},\ \bibinfo
  {pages} {13} (\bibinfo {year} {1997})}\BibitemShut {NoStop}%
\bibitem [{\citenamefont {Sato}(2000)}]{Sato2000}%
  \BibitemOpen
  \bibfield  {author} {\bibinfo {author} {\bibfnamefont {S. Sato {\it et al.}}},\ }\href {\doibase 10.1364/AO.39.004616}
  {\bibfield  {journal} {\bibinfo  {journal} {Applied optics}\ }\textbf
  {\bibinfo {volume} {39}},\ \bibinfo {pages} {4616} (\bibinfo {year}
  {2000})}\BibitemShut {NoStop}%
    \bibitem [{\citenamefont {Michimura}(2018)}]{Michimura2018}%
  \BibitemOpen
  \bibfield  {author} {\bibinfo {author} {\bibfnamefont {Y. Michimura {\it et al.}}},\ }{\bibfield
  {journal} \ \Eprint {https://arxiv.org/abs/1804.09894v2} {arXiv:1804.09894}}
  \BibitemShut {NoStop}%
  \bibitem [{\citenamefont {Kimble}(2001)}]{Kimble2001}%
  \BibitemOpen
  \bibfield  {author} {\bibinfo {author} {\bibfnamefont {H. J. Kimble, Y. Levin, A. B. Matsko, K. S. Thorne and S. P. Vyatchanin}},\ } {\bibfield  {journal} {\bibinfo  {journal}  {Physical Review D}\ }\textbf {\bibinfo {volume} {65}},\ \bibinfo {pages} {022002}
  (\bibinfo {year} {2001})}\BibitemShut {NoStop}%
  \bibitem [{\citenamefont {Buonanno}(2001)}]{Buonanno2001}%
  \BibitemOpen
  \bibfield  {author} {\bibinfo {author} {\bibfnamefont {A.Buonanno and Y.Chen}},\ }\href {\doibase 10.1103/PhysRevD.64.042006}{\bibfield  {journal} {\bibinfo  {journal}  {Physical Review D}\ }\textbf {\bibinfo {volume} {65}},\ \bibinfo {pages} {022002}
  (\bibinfo {year} {2001})}\BibitemShut {NoStop}%  
  \bibitem [{\citenamefont {Evans}(2008)}]{Evans2008}%
  \BibitemOpen
  \bibfield  {author} {\bibinfo {author} {\bibfnamefont {M. Evans, S. Ballmer, M. Fejer, P. Fritschel, G. Harry, and G. Ogin}},\ } \href {\doibase  10.1103/PhysRevD.78.102003}
  {\bibfield  {journal} {\bibinfo  {journal}  {Physical Review D}\ }\textbf {\bibinfo {volume} {78}},\ \bibinfo {pages} {102003}
  (\bibinfo {year} {2008})}\BibitemShut {NoStop}%
  \end{thebibliography}

%\begin{thebibliography}{9}%
%\makeatletter
%\let\auto@bib@innerbib\@empty
%\end{thebibliography}%

\end{document}